\documentclass[12pt,preprint]{aastex}
\shorttitle{Nova Shell of T Pyx}
\shortauthors{Schaefer, Pagnotta, \& Shara}
\begin{document}

\title{The Nova Shell and Evolution of the Recurrent Nova T Pyxidis}
\author{Bradley E. Schaefer, Ashley Pagnotta}
\affil{Physics and Astronomy, Louisiana State University,
    Baton Rouge, LA, 70803, USA; schaefer@lsu.edu}
\author{Michael M. Shara}
\affil{Department of Astrophysics, Division of Physical Sciences, American Museum of Natural History, Central Park West at 79th Street, New York, NY 10024}

\begin{abstract}

T Pyxidis is the prototypical recurrent nova (RN), with five eruptions from 1890 to 1967 and a mysterious nova shell.  We report new observations of the nova shell with the {\it Hubble Space Telescope} ({\it HST}) in the year 2007, which provides a long time baseline to compare with {\it HST} images from 1994 and 1995.  We find that the knots in the nova shell are expanding with velocities ranging from roughly 500-715 km s$^{-1}$, assuming a distance of 3,500 pc.  The fractional expansion of the knots is constant, which implies no significant deceleration of the knots, which must have been ejected by an eruption close to the year 1866.  We see knots that have turned on after 1995; this shows that the knots are powered by shocks from the collision of the `1866' ejecta with fast ejecta from later RN eruptions.  We derive that the `1866' ejecta has a total mass of $\sim10^{-4.5}$ M$_{\odot}$, which with the low ejection velocity shows that the `1866' event was an ordinary nova eruption, not an RN eruption.  This also implies that the accretion rate before the ordinary nova event must have been low (around the $4\times10^{-11}$ M$_{\odot}$ yr$^{-1}$ expected for gravitational radiation alone), and that the matter accumulated on the surface of the white dwarf for $\sim$750,000 years.  The current accretion rate ($>10^{-8}$ M$_{\odot}$ yr$^{-1}$) is $\sim$1000$\times$ higher than expected for a system below the period gap, with the plausible reason being that the `1866' event started a continuing supersoft source that drives the accretion.  The accretion rate has been declining since before the 1890 eruption, with the current rate being only 3\% of its earlier value.  The decline in the observed accretion rate shows that the supersoft source is not self-sustaining; we calculate that the accretion in T Pyx will effectively stop in upcoming decades.  With this, T Pyx will enter a state of hibernation lasting for an estimated 2,600,000 years before gravitational radiation brings the system into contact again.  Thus, T Pyx has an evolutionary cycle going from an ordinary CV state (lasting 750,000 years), to its current RN state (lasting little longer than a century), to a future hibernation state (lasting 2,600,000 million years), and then repeating this cycle.

\end{abstract}

\section{Introduction}

	Cataclysmic variables (CVs) with multiple nova outbursts in a given century are known as
recurrent novae (RNe). The dramatic outbursts are the product of thermonuclear runaway
in the accreted material on the surface of the white dwarf (WD). The outburst mechanism
is identical to that of classical novae (CNe), but the presence of a high-mass WD
($M_{WD}\gtrsim 1.3$ M$_{\odot}$) and a high accretion rate lead to the short recurrence 
time scale.  With these required conditions, RNe are strong candidate Type Ia supernova progenitors.  Only ten RNe are known in our galaxy; the prototype RN is T Pyxidis. 

	T Pyx was the first known RN, discovered by H. Leavitt in 1913 (Pickering 1913) when she found the 1902 eruption on the Harvard plates and searched through older plates to find the 1890 eruption.  Eruptions of T Pyx have been observed in 1890, 1902, 1920, 1944, and 1967, giving an
apparent recurrence time scale of approximately 20 years (Schaefer 2009).  This time scale, however, predicts an eruption around 1987, which did not occur. The five eruptions have identical light curves, each with a peak magnitude of 6.4 in the visual and a $t_3$ of 62 days, where $t_3$ is the amount of time it takes for the brightness to decline by three magnitudes from peak (Schaefer 2009).  The quiescent B-band magnitude of the system has fallen systematically from 13.8 mag before the 1890 eruption, to 15.5 mag in 2004 (Schaefer 2005), and now to 15.7 mag in 2009.  The orbital period of T Pyx was discovered to be {\it below} the period gap at 1.83 hours (Schaefer et al. 1992) with extensive photometric (Patterson et al. 1998) and spectroscopic confirmation (Uthas 2009).  With this, it is surprising that all measures of the accretion rate (e.g., Patterson et al. 1998; Selvelli et al. 2008) are at the very high value of $>10^{-8}$ M$_{\odot}$ yr$^{-1}$.
	
	T Pyx is surrounded by a unique nova shell. This shell, with a radius of $\sim5$ arc-second, was discovered by Duerbeck \& Seitter (1979) using the ESO 3.6m telescope.  Williams (1982) used the CTIO 4m to image T Pyx in H$\alpha$+[N II], and found the shell to be spectrally similar to planetary nebulae, and to have close to solar abundances.  The obvious presumption was that the shell had been ejected during the 1967 eruption.  Shara et al. (1989) compared images from 1980 and 1985 to find that the shell was apparently {\it not} expanding, and could certainly not have come from the 1967 eruption.  They also discovered a faint outer halo extending out to radius $\sim10$ arc-seconds.  Contini \& Prialnik (1997) presented a model for the T Pyx shell that successfully reproduces the spectral line fluxes of the shell by shock heating when the fast ejecta from one nova eruption run into the slower ejecta from an earlier eruption.  {\it Hubble Space Telescope} ({\it HST}) images from 1994 and 1995 revealed the shell to be composed of over 2000 unresolved knots (Shara et al. 1997).  Mysteriously, the knots were {\it not} seen to be expanding radially, although the time baseline was fairly short.  The origin and nature of the nova shell are essentially unknown, with many proposed possibilities.
	
	T Pyx might be a well-observed system, but it still has many mysteries.  First, why didn't it erupt around the year 1987, as appropriate for its prior recurrence time scale?  Second, its orbital period places it below the period gap and hence it `should' have a low accretion rate driven by gravitational radiation loss, so why is its observed accretion rate over a thousand times higher?  Third, why is the nova shell so unlike almost all other nova shells, being composed of a myriad of small knots?  Fourth, why is this shell seen to be {\it not} expanding?  Fifth, why has T Pyx been suffering a unique secular decline in quiescent brightness by 1.9 magnitudes over the last 120 years?  Sixth, will T Pyx ultimately become a supernova?  We present answers for each of these mysteries.
	
\section{HST Observations}

	The previous {\it HST} images of T Pyx had a short time baseline, from 1994.1 to 1995.8, and the knots were seen to not have any significant radial expansion (Shara et al. 1997).  Over a decade later, we use another set of {\it HST} images to look for the expansion velocity with much greater sensitivity due to the longer baseline.  Such images can also look for brightness changes (to constrain the power source of the knot emission) and for the presence of an inner shell which has expanded to a far enough distance from the bright central star to be recognizable.  We were granted {\it HST} time in 2007 with the ACS instrument.  The greater resolution and sensitivity of ACS (as compared to WFPC2) would allow for a deeper image than the 1994/5 images in just one orbit.  Unfortunately, the ACS detector was disabled before our observations, so our program was rescheduled for WFPC2.  The primary science loss was in the sensitivity; our observations did not go very deep.  Nevertheless, we obtained an image which shows 30 knots with good significance, which is adequate for our primary science goals.  A benefit of this situation is that our later observations used exactly the same detector/filter/telescope combination as the earlier observations, which is reassuring for our comparisons.

	The primary input for this paper is this new set of {\it HST} images.  We chose to use the F658N filter because this is the most sensitive to the knots (which are bright in $H\alpha$) and because most of the exposures in 1994 and 1995 were through the same filter.  The F658N filter has a central wavelength of 659.1 nm and a bandwidth of 2.9 nm, which covers not only the $H\alpha$ line but also the bright [NII] line (Williams 1982).  We chose to have the images centered slightly towards the nearby star, labeled as `Check' in Schaefer (2009) and as `155' by the {\it AAVSO}, which is 12 arc-seconds directly to the south.  Check provides a photometric standard (B=16.56, V=15.81, R=15.33, I=14.95, J=14.31, H=13.94, and K=13.89; Schaefer 2009) and a point-spread-function (PSF) standard.  We obtained two orbits of images, each with a total exposure time of 1900 seconds (for a total of 3800 seconds) on 2007 June 29.  Each of the two images were broken up into the usual pair for elimination of cosmic rays.  These images were processed and combined with the standard {\it HST} pipeline, drizzled using the PyRAF {\it multidrizzle} task, and then combined via IRAF's {\it  imcombine} task.  Everything of interest fits into the field of the Planetary Camera on WFPC2.  The field of view is 800$\times$800 pixels, which is 35$\times$35 arc-seconds.  The nominal pixel size is 0.046 arc-seconds.
	
	Our full, final {\it HST} image from 2007 is displayed in Figure 1.  The 2007 image shows T Pyx just north of center, the Check star near the bottom edge, three faint background stars, and many knots in a roughly symmetrical shell around T Pyx.  Detailed photometry also shows an apparently smooth source above background that is centered on T Pyx, due to unresolved faint knots and possibly a smooth shell.  Many knots are readily identified, and all of these are unresolved.  The majority of these knots appear 100-130 pixels (4.6-6.0 arc-seconds) from T Pyx, while the remainder are mostly 70-100 pixels (3.2-4.6 arc-seconds) from T Pyx.  The knots are shown more clearly in Figure 2.
	
	We have compiled a list of the positions and magnitudes of knots from our 2007 image, presented in Table 1.  We have taken a flux limited sample of 30 knots.  Fainter knots are certainly visible, but their measured properties and existence are increasingly less confident as the knots get fainter than our threshold.  Each knot is designated with a prefix `K' and a running number, in order of radial distance from T Pyx.  We have used IRAF routines to measure the centroid position and aperture photometry (with a 2.0 pixel radius circular aperture) to measure the brightness of each knot.  We also performed centroid positioning and aperture photometry on T Pyx and Check.  Each knot position is reported relative to that of T Pyx in units of pixels, with the positive X-axis towards the west and the positive Y-axis towards the north (columns 2 and 3, in pixels).  The one-sigma error bar on these positions is 0.2 pixels as determined by the consistency between measures on our two orbits.  We also report the calculated radial distance ($R_{2007}$, in pixels) and the position angle (measured westward from north in degrees) from T Pyx (see columns 4 and 5).  Each knot magnitude is reported relative to that of Check ($\Delta m_{07}$ in magnitudes), as given in column 6.  The one-sigma error bar on these magnitudes is 0.10 mag as determined by the consistency between measures on our two orbits.

	In the time since 1995, it is possible that knots or expanding shells might appear close in to T Pyx.  In Figures 1-3, T Pyx itself darkens the images within 10 pixels of the center.  Knots or shells interior to this can be detected by subtracting out the PSF of Check, scaled to the flux of T Pyx before subtraction. The subtracted image showed differences at the part-per-thousand level which is consistent with normal and irreducible variations in the PSF.  We do not see any systematic structure corresponding to a `shell' centered on T Pyx, or any significant residuals with a PSF shape of their own, which is to say that we do not see any knots close in to T Pyx.
	
	The primary science here comes from a detailed comparison between our image from 2007.5 and the {\it HST} images from 1994.1 and 1995.8.  The earlier images are described fully in Shara et al. (1997).  We have extracted the original WFPC2 images and reprocessed them using the same technique we used on the 2007.5 data.  (The old images remain identical to those displayed in the earlier paper, but we repeated the processing  primarily as a precaution.)  The relative positions of T Pyx and the Check star are constant to within a quarter of a pixel on all images and sub-images.
	
	A casual comparison of the 1995.8 and 2007.5 images shows many important points.  First, the 1995.8 image goes much deeper and shows many more knots.  This is expected given the relative exposure times.  Nevertheless, the 2007.5 exposure is adequate to see many knots, which is what we need.  Second, the overall structure of the knots is the same.  Third, the individual knots form random patterns that appear in {\it both} the 1995.8 and 2007.5 images.  This is important as the repetition of such patterns would be essentially impossible unless the same knots appear in both images.  We can therefore track the motion of specific knots and monitor changes in their brightness.  Fourth, a few knots appear on the 2007.5 image at a position which is completely empty in the 1995.8 image.  Something has caused these knots to light up between 1995.8 and 2007.5; this clue will determine the mechanism by which the knots shine.  Fifth, some knots appear to be somewhat fainter in 2007.5 than in 1995.8.  This fading provides a measure of the electron density in the knots.
	
	With aperture photometry, we can compare the brightness of T Pyx with that of Check for our F658N images.  On the first orbit, T Pyx was 0.10 mag brighter than Check, while on the second orbit it was 0.21 mag brighter.  This variation is typical of T Pyx, arising from the usual flickering and orbital modulation of the system (Schaefer 1990; Schaefer et al. 1992; Patterson et al. 1998).
	
	We also obtained quasi-simultaneous photometry with the 1.3-m SMARTS telescope on Cerro Tololo in Chile.  These observations consist of BVRI magnitudes on six consecutive nights from 2007 June 21-26.  We found average magnitudes of B=15.65, V=15.56, R=15.38, and I=15.17 for T Pyx.  The RMS scatter in these magnitudes is 0.05 mag.

\section{Expansion of the Nova Shell}

	A quick measure of the positions of knots prominent in 1995 and 2007 shows that the knots are expanding radially.  A visual example of this is shown in Figure 3, where positions of the 1995 and 2007 images are presented with identical scales, positioning (relative to T Pyx), and angular sizes.  T Pyx itself is vertically lined up in both panels, so that any radial expansion towards the right is readily seen in the misalignment of the knots.  With this, the easily recognizable `constellation' of bright knots slightly right of center is clearly seen to have a substantially and significantly larger radial distance in 2007 than in 1995.  This demonstrates that the T Pyx shell is expanding.  Also, as we look from left to right, the positional differences become larger.  This demonstrates that the expansion of the shell is homologous, where the increase in radial distance is proportional to the radial distance.
	
	Our conclusion (homologous expansion) can in principle be compromised by the selection of knots in the two images.  There are three ways that this could happen.  First, we could select knots from the 1995 image based on some biased idea of where the knots {\it should} be, and the selected knots would then confirm the bias.  Second, the presence of many knots in the 1995 image might provide too many candidate knots for assignment with a knot in the 2007 image, which could cause a possibly large systematic error.  Third, the knots are seen to vary in brightness, so a `Christmas tree light' situation could result in a pairing of knots on images that are only neighbors.  Fortunately, the density of knots is sufficiently low that these potential problems do not seem significant, and the distinction of the pattern of knots (i.e., their `constellations') is clear, so only causal connections between 2007 and 1995 are made.
	
	To provide a formal and unbiased test of our knot identification, we have performed a cross correlation analysis between the 2007, 1995, and 1994 images.  For this, we have taken each knot position in 2007 and constructed a series of positions where the radial distance from T Pyx is increased by a factor $F$ (for $0.80\leq F \leq 1.20$).  For example, if we had a knot with a 2007 position exactly 100 pixels north of T Pyx, then our constructed series of positions would be 80, 81, 82,..., 118, 119, and 120 pixels north of T Pyx.  With the 1995 image, we placed a 1.0 arc-second radius photometry aperture at each of the positions in the series and measured the total flux inside the aperture.  We also measured the flux inside a similar aperture at the correct center of the knot on the 2007 image.  The cross correlation for each knot was then a product of the knot flux in the 2007 aperture times the flux inside each of the 1995 apertures.  If there was no expansion from 1995 to 2007, then all the cross correlations would have a peak at $F=1.00$.  If the shell has a homologous expansion, then all the cross correlations should have a peak at the same $F$ value.  With this, we have a tool for a quantitative and unbiased measure of the shift of each knot.
	
	The cross correlations for five representative knots are given in Figure 4, where the product of the fluxes is given as a function of $F$.  The cross correlations have been normalized by their peak value.  Some of the knots have multiple peaks, with one of the peaks being caused by the random coincidence of another knot nearby.  However, most knots have only one peak, and this peak is close to $F=0.91$.  The cross correlation summed over all 30 knots is presented in Figure 5.  We see a single and highly-significant peak around $F=0.91$.  This result guarantees that the connections between knots in the 1995 and 2007 images are correctly done.
	
	In Figure 5, we have also shown the summed cross correlation for the 1994 image.  We again see a single and highly-significant peak close to $F=0.91$. This provides the quantitative and unbiased evidence of the homologous expansion of the T Pyx shell.  We can now quantify the expansion of the shell.  We measure the expansion $F$ by four different methods.
	
	The first method is to take the best value from Figure 5.  The peaks are at $F=0.912$ and $F=0.907$ for 2007.5-1995.8 and 2007.5-1994.1, respectively.  The uncertainties will be close to the half width at half maximum of the peak divided by the square root of the number of knots.  With this, we find $F=0.912\pm0.004$ and $F=0.907\pm 0.005$ for the two offsets.
	
	The second method makes use of the peak $F$ values for all knots.  For example, in Figure 4, the peaks are at 0.906, 0.906, 0.925, 0.930, and 0.933.  The average and the uncertainty on the average will give the best $F$ value.  This method is different from the first method primarily in how it weights the individual knots, in this case with knots being of equal weight as opposed to weighting by brightness.  We get $F=0.917\pm0.003$ for 2007.5-1995.8 and $F=0.905\pm 0.003$ for 2007.5-1994.1.
	
	The third method uses the radial shifts measured for each of the knots.  We have measured the change in radial position, based on the centroid position, and performed standard aperture photometry for each of the knots in the 2007 image that have corresponding knot images in the 1994 or 1995 images.  These measures are tabulated in Table 1, where columns 6-8 and 9-11 give the position angle (in degrees), radial distance (in pixels), and differential magnitude (in mags) for each knot for 1995 and 1994 respectively. The next two columns list the distance ratios $R_{1994}/R_{2007}$ and $R_{1995}/R_{2007}$, and the last two columns list the differences in magnitudes from 2007 to 1995 and 2007 to 1994.  The average of these distance ratios will be $F$, while the uncertainty on this average will be the uncertainty on $F$.  Thus, $F=0.919\pm0.003$ for 2007.5-1995.8, and $F=0.902\pm 0.006$ for 2007.5-1994.1.
	
	The fourth method uses the slope of the change in radial positions of the knots as a function of the angular distance from T Pyx.  In Figure 6, we have constructed such a plot for the changes in radial distances ($\Delta R$) from 2007 to both of the earlier images.  We see that the two sets of differences lie along a fairly well-defined line that passes through the origin.  The $\Delta R$ values for 2007.5-1994.1 are systematically larger than for 2007.5-1995.8, and this shows that the shell expanded somewhat over the short time baseline from 1994.1-1995.8.  Simple linear fits show the intercept to be within a fraction of a pixel of zero; this implies that the angular rate of expansion is proportional to the radius, so the expansion is homologous.  The $F$ values are simply one minus the slope, or $F=0.918\pm0.005$ for 2007.5-1995.8, and $F=0.913\pm 0.006$ for 2007.5-1994.1.
	
	We have now measured the expansion by four different means.  The scatter in these values is likely a combination of both the measurement errors and the differing weights and methods.  The RMS scatter of the four values is similar to the average quoted uncertainties.  So we adopt the average $F$ values, and use the RMS scatter of the four values as the error bar in $F$.  Thus, we conclude $F=0.917\pm0.003$ for 2007.5-1995.8 and $F=0.907\pm 0.005$ for 2007.5-1994.1.
	
	We can now calculate the expansion factor for the shell.  That depends on the distance to T Pyx, which is poorly known.  Schaefer (2009), Patterson et al. (1998), Webbink et al. (1987), and Selvelli et al.  (2008) discuss the poor constraints, all coming to the conclusion that the best estimate for the distance is close to 3500 pc.  The uncertainties are given to have ranges of $\pm1000$, $\pm1000$, $\pm1725$, and $\pm350$ pc, respectively.  (The smallest value is not realistic.  It is an artifact of the method the authors used to derive the uncertainty, which considers the scatter of five different estimates of which four are from essentially identical methods and input.  This procedure forces a too-small error estimate.)  We adopt a distance of $3500\pm1000$ pc.  For convenience, we define $d_{3500}$ to be the distance to T Pyx divided by 3500 pc.  A motion of one pixel (0.046 arc-seconds) from 1995.8 to 2007.5 corresponds to $65~d_{3500}$ km s$^{-1}$ of transverse velocity.  The outermost knots (at $R_{2007}=130$ pixels, or 6.0 arc-seconds, or 0.102 pc) presumably correspond to the case of transverse motion and maximum velocity, and for these we have an average $\Delta R = 11\pm1$ pixels.  Thus the maximum velocity of the knots is $V_{max}=715\pm65 ~d_{3500}$ km s$^{-1}$.  
	
	The average velocity is harder to determine because non-transverse motion and sub-maximal velocities can be confused.  To get an estimate of the range of velocities, we have constructed a Monte Carlo simulation of knots being ejected (with no deceleration) in random directions with respect to the line of sight and with a range of velocities uniformly distributed between $V_{min}$ and $V_{max}$.  We seek to determine what value of $V_{min}$ reproduces the observed radial distribution of knots.  For example, the possibility of of $V_{min}$ being close to 715 km s$^{-1}$ is rejected because then the predicted radial distribution would show a cusp at 130 pixels.  Also, for small values of $V_{min}$, the fraction of knots less than 70 pixels from T Pyx would be much higher than observed.  With 30 knots in 2007.5, we can only have moderate accuracy in the derived velocity distribution.  This can be overcome by using the radial distribution from the 1995 image (see Fig. 13 of Shara et al. 1997). In all cases, the systematic problems (whether the real velocity distribution is just a constant over a given range, and whether the knot distribution is the idealized spherical distribution or is asymmetric, perhaps bipolar) are likely to dominate over the counting statistics.  With this, we get $V_{min}=500\pm50$ km s$^{-1}$.  In all, our observations show that the knots have an expansion velocity that varies over the range 500-715 km s$^{-1}$), for which a reasonable average is 600 km s$^{-1}$.
	
	This average velocity is substantially different from the maximum velocity of 2000 km s$^{-1}$ as observed from spectral line profiles during the 1967 eruption (Catchpole 1969).  The difference will be resolved as we realize that the eruption that emitted the visible knots have substantially different properties (including expansion velocity) than do the RN eruptions like in 1967.  The observed average knot velocity of 600 km s$^{-1}$ sets the scale for collisions between these knots and any other gas in the system, with the relative velocity of collision being perhaps somewhat larger or smaller depending on the circumstances.  The kinetic energy per particle corresponds to energy in the X-ray band, so we could plausibly expect some X-ray emission caused by collisions.  Indeed, in 2006, hard X-ray emission was seen by {\it XMM}, with the cause attributed to shock heated ejecta (Selvelli et al. 2008).

	With this measured expansion rate for the nova shell, we can now perform the simple calculation of extrapolating this motion back to the time when the knots were ejected from T Pyx.  For a first epoch image that is $\Delta Y$ years before our current image, the year of eruption is $2007.5-\Delta Y(1-F)^{-1}$.  The one-sigma uncertainty is $\sigma_F \Delta Y (1-F)^{-2}$.  For the two first epochs, we get $1867\pm5$ and $1864\pm7$.    These two values share half of their data (i.e., the 2007 image), so they cannot be simply combined.  We take as our final value for this as $1866\pm5$.  This date will give the real year of the eruption only if there is no significant deceleration or acceleration of the knots.  As we will demonstrate in the next section, this is exactly the case.

\section{Deceleration Models}

	The knots ejected by T Pyx will undergo deceleration to some degree due to surrounding gas, either from the ordinary interstellar medium (ISM) or from previous nova shells.  Several plausible cases can be proposed for the deceleration, and we examine three specific scenarios which span the possibilities.
	
	Unfortunately, all we really know is the current expansion velocity ($\sim 600~d_{3500}$ km s$^{-1}$, with an uncertainty due to the poorly-known distance) and the maximum expansion velocity from the 1967 eruption ($2000$ km s$^{-1}$, Catchpole 1969).  The velocity from Catchpole corresponds to the half-width at zero-intensity for the $H\beta$ line, and thus the maximal expansion velocity.  This velocity is independent of distance, corresponds to the speed at the photosphere along the line of sight, and presumably measures the expansion velocity all around the shell at some given distance.  As this velocity is nearly constant over the entire eruption, during which the photosphere spans a wide range of distances, the bulk of the ejected material likely has velocity comparable to 2000 km s$^{-1}$.  Catchpole's velocity was only measured for the 1967 eruption, but all of T Pyx's eruption light curves are identical (Schaefer 2009) and the triggering conditions do not change (Schaefer 2005), so it seems likely that the expansion velocity in each of the observed RN eruptions was identical to that measured by Catchpole.   This is the basic information that an acceptable deceleration model must be able to reproduce.
	
	The model results are presented graphically in Figure 7, which plots the radial distance of knots as a function of the year.  In this graph, the curves rise out of the  horizontal axis on the year of the eruption, with a slope that gives the initial expansion velocity.  Deceleration is represented by the curve flattening as time increases.  The slope is the velocity at that time.  The slope during the years from 1995.8 to 2007.5 should be close to 600 km s$^{-1}$ since that is the currently-observed expansion velocity.  Each of the deceleration models below is represented by one curve in the figure.
	
	\subsection{No Deceleration}
	
	We first consider the scenario in which the knots we currently see expanding have suffered no significant deceleration (or acceleration).  This is the case for the knots arising from a big classical nova event in the 1800s (the preferred case, as concluded later in this paper).  In this case, there would have been no contribution to the surrounding interstellar medium from previous eruptions, so the background medium would be too thin to cause significant slowing.  This big eruption must be completely different from the later RN events and thus need not have the initial expansion velocity of 2000 km s$^{-1}$.
	
	The simple no-deceleration model has an unambiguous eruption date.  From the previous section, we found that the originating explosion had a date of $1866\pm5$ for the case of no significant deceleration.  It is important to note that this date is completely independent of the distance to T Pyx.  This corresponds in time to an eruption perhaps two cycles before the first known eruption in 1890 which was completely different than the 1890 eruption.  This model is represented in Figure 7 by the straight line that rises in the year 1866.
	
	\subsection{Three Models With Deceleration}
	
	As the knots expand, they can run into slower gas, with this coming from either the interstellar medium or from prior eruptions.  As a knot moves through this slower gas, it will act as a snowplow and gain in mass.  As the individual atoms interact with the knots, they will rapidly be thermalized and become part of the knot, while collectively a shock will be sent through the knot.  The combined specific momentum (for the knot plus the swept-up gas) will decrease, so the knot will slow down.  This deceleration can be observed in the knot's expansion from the central star.
	
	The possibility of deceleration is suggested by three points.  First, the observed expansion velocity (based on the motion of the knots) is $\sim 600$ km s$^{-1}$ for the years 1995-2007, whereas the observed expansion velocity (based on the HWHM of spectral lines from the last eruption) is $\sim 2000$ km s$^{-1}$ in 1967.  The stark difference between these velocities implies a deceleration, {\it if} the two velocities are referring to the same gas.  Second, {\it if} the knots come from the 1967 eruption, then they would have to expand with a velocity of near $\sim 2000$ km s$^{-1}$ to get out to their current radial position, yet they are now expanding with a velocity of $\sim 600$ km s$^{-1}$ and this implies that there must be a deceleration.  Third, there must be some amount of gas being swept-up by the expanding knots, and this can only decelerate the knots to some extent.
	
	We have constructed a kinematic model of this deceleration.  The primary equation for our model is simply the conservation of momentum as the knots sweep up slower moving material.  The velocity of the swept-up gas will depend on the scenario, where gas from the interstellar medium is assumed to have zero velocity with respect to the central star, while material from a prior eruption is taken to have a velocity corresponding to the time since the prior eruption and the radial distance from the central star.  The density of the swept up gas will also depend on the scenario, with the interstellar medium assumed to have a constant density.  The density of the gas from a prior eruption will depend on the velocity distribution of the prior ejecta, and we have adopted a uniform distribution between some minimum and maximum velocities ($V_{min}$ and $V_{max}$).  If the velocity range is narrow then the knots will suddenly run into a shell, while if the velocity range is wide then the knots will be steadily plowing through an expanding medium.  The density of the prior ejecta will be falling off as the inverse-square of the radial distance.  The knots themselves are taken to have some cross section (which can either be constant or increase by simple radial motion) for sweeping up material and some mass (which increases as material is swept up).  With this, it is simple to construct a kinematical model that derives the radial distance and velocity of each knot.  For some scenarios this can be solved analytically, while for other scenarios we have resorted to numerical integration.
	
	Our kinematic model has limitations.  We have not attempted any hydrodynamical solutions nor calculated the details of any shocks within the knots.  This is fine, as we cannot see any shocks and the knots are all unresolved.  A traditional approach would be to apply the Sedov solution (e.g., Chevalier 1982), as was done for T Pyx within a completely different scenario by Contini \& Prialnik (1997).  But the Sedov solution does not apply to the scenarios we are considering because the expansion velocity of the prior ejecta is not constant, and also because we are considering discrete knots instead of some expanding shell with an $r^{-7}$ density structure.  Only the knots are visible, and these behave as discrete entities gaining mass and slowing down.
	
	For models with deceleration, we will examine three scenarios.  The first scenario is where the knots expand without significant deceleration until they run into and coalesce with substantially slower ejecta from an earlier eruption.  This case is plausible because the ejecta mass appears to be predominantly in the form of knots, and thus ejecta from a later eruption would be slowed down when one of the knots impacted a knot from an earlier eruption.  The second scenario is that the deceleration is caused by relatively slow-moving gas ejected by the previous nova eruption.  For example, the fast moving knots from the 1967 eruption will fairly quickly catch up with (and be slowed by) the slow-moving gas from the 1944 eruption.  The third scenario has the ejecta being slowed by interactions with the interstellar medium.  This situation could apply to the ejecta from a big eruption in the middle 1800s.  It is possible that the ejection velocity from a nova event on a massive, cool white dwarf might be much larger than the current expansion velocity of the knots (for the RN events), and thus there might be some deceleration, which could only come from the uniform ISM.
		
	Our first scenario with deceleration is that the outgoing knots are suddenly slowed down due to collisions with knots or sharp edged shells ejected in earlier eruptions.  If the knot ejection happened during the last eruption (in 1967), then a small problem with this model is that the required initial velocity must be $\gtrsim2700$  km s$^{-1}$ for the radial distance to match the observations in 1995.8.  This is not a significant problem, as either the distance to T Pyx could be smaller than the adopted value of 3,500 pc, or the initial expansion velocity could be somewhat larger than the Catchpole velocity.  Upon reaching the required radial distance in the year 1995, the fast-moving knot can slow to the required velocity by colliding with a slow-moving knot or shell.  To take a specific example, a collision with a knot whose velocity is 300 km s$^{-1}$ that is 7.0 times more massive will produce the observed knot velocity.  The expansion history within this model is represented in Figure 7 by a singly-broken line rising up from the year 1967, with one segment lying on top of the other models for the time interval 1995.8-2007.5.
	
	A substantial problem with this model is that it cannot account for all of the observed knots going outward with a final velocity $\sim$600 km s$^{-1}$.  This problem arises from the conservation equation.  We can imagine scenarios where initial velocities of the knots and the impacted gas are both held constant, but the knots all have widely varying masses (see Shara et al. 1997) with the result being that the final velocities will all be widely different.  For example, for initial velocities of 2000 km s$^{-1}$ and 300 km s$^{-1}$, the final velocity changes from 320 to 1150 to 1980 km s$^{-1}$ as the mass ratio changes from 0.01 to 1.0 to 100.  That is, for the observed broad distribution of knot masses, there is no way that the observed expansion velocities would be nearly constant.  As such, the sudden deceleration model cannot be accepted.
	
	Our second scenario is that the deceleration is caused by relatively slow-moving gas ejected by the previous nova eruption.  For example, the fast-moving knots from the 1967 eruption (moving at $V_{max}$) will fairly quickly catch up with (and be slowed by) the slow-moving gas from the 1944 eruption (moving at $V_{min}$).
	
	Within the second scenario, the knots seen around T Pyx could not have come from the 1967 eruption.  Most of the deceleration would have to occur soon after the time when the knots hit the prior ejecta.  To get the knot out to a large distance before the deceleration, we must have a near-maximal $V_{min}$; then the collision with the inner edge of the 1944 shell must be at low relative velocity and little deceleration can result.  As such, within this model, we can disprove the idea that the observed knots are from the 1967 eruption.
	
	The same problem holds for earlier eruptions.  The prior nova shell has its highest density at its inner edge, which is also the place where the velocity difference is the largest, so most of the slowdown in the knots occurs close to the time when the outgoing knots hit the inner edge.  Again we have the case of a fairly sudden deceleration.  To match the observed 1995.8-2007.5 velocity, the sudden deceleration will always have to lower the initial velocity down to roughly 600 km s$^{-1}$.  In the plot of Figure 7, this will always look like a steeply sloped line rising for one of the eruptions between 1890 and 1944, which will then make a fairly sharp turn so as to parallel the topmost straight line in the figure.  We can see that the deceleration must occur at a relatively large radial distance, and this can only occur if the minimum velocity of the previous eruption is large.  In addition, the moving knot will rapidly slow down to the minimum velocity of the previous shell and simply be carried along, so that its apparent motion from 1995.8-2007.5 will either be too fast or at too small a radial distance.  With the resultant small velocity difference, to get the required deceleration to near 600 km s$^{-1}$, the masses in the knot and/or the prior nova shell must become implausibly extreme.
	
	Figure 7 shows a typical best case.  Here, we have assumed that the knots come from the 1902 eruption with initial velocity 2000 km s$^{-1}$ running into the slow-moving ejecta with 670 km s$^{-1}$ from the 1890 eruption.  The knots will hit the earlier ejecta in the year 1908.0 at a radial distance of 0.012 pc.  The radial distance in the year 2007.5 is reproduced when the knots have the maximal size of 0.001 pc and a mass of $10^{-11}$ M$_{\odot}$.  With this, the knot velocity drops to 1000 km s$^{-1}$ in a quarter of a year after it hits the inner edge of the shell of the previous eruption.  The 1995.8-2007.5 average velocity is 685 km s$^{-1}$, which is close enough to our observed value to be acceptable.  A substantial problem is that this situation has the density in the knot equal to 0.1 hydrogen atoms per cubic centimeter, whereas the observed value is $\gtrsim10^4$ times larger (Shara et al. 1997).  Also, the mass of the knot is much smaller than that of the observed knots (Shara et al. 1997).  Attempts to solve these problems by increasing the density and mass of the knot only make for higher inertial mass and hence unacceptably high terminal velocity.  Alternatively, we can get essentially the same plot with the more reasonable values of mass of $10^{-9}$ M$_{\odot}$ and density of $10^4$ atoms per cubic centimeter by reducing the knot radius to 0.0001 pc, but the cost then is that $M_{ejecta}$ has to be increased to the unreasonable level of 0.01 M$_{\odot}$ to get enough deceleration for a knot with such a small cross section.  In all, there is no reasonable solution for any eruption date within this model.
	
	Another critical problem with this deceleration model is that the terminal velocity (as observed from 1995.8-2007.5) depends greatly on the mass of the knot.  For example, in the case of the last paragraph, an increase to $10^{-10}$ M$_{\odot}$ causes the terminal velocity to increase to 940 km s$^{-1}$.  We know that the knots vary in mass by almost two orders of magnitude (Shara et al. 1997).  Yet all the observed knots are expanding at the same fractional rate.  This would only be possible (within this  model) if all the knots had nearly the same mass, and this seems certainly wrong given the wide range in knot brightnesses.  Thus, for a second strong reason, we see that our second scenario is not acceptable.
	
	Our third scenario with deceleration has the ejecta being slowed by interactions with the interstellar medium.  Here, the density of the medium is presumed to be constant and to have an average velocity of zero.  Within this model, we find that the knots could plausibly come from the big $\sim$1866 eruption, or from the 1890, 1902, 1920, or 1944 RN eruptions.  (The knots cannot have come from the 1967 eruption for any plausible conditions.)  For appropriate assumptions of initial velocity, the knot mass, the knot radius,  and the density of the interstellar medium, we can find situations that match the observations.  The model for 1920 is shown in Figure 7; it is constructed with  an initial velocity of 2000 km s$^{-1}$, a knot mass of $10^{-8}$ M$_{\odot}$, a knot radius of 0.0805 pc in 1995.8 and 0.0877 pc in 2007.5, and ISM density of 3 atoms per cubic centimeter.
	
	Although there are feasible solutions for individual knots within this model, it has the same fault as the previous scenarios with deceleration; the velocity history depends strongly on the knot mass, and since the masses vary greatly, we inevitably end up with knots having greatly different velocities during the 1995.8-2007.5 time period, which contradicts the observations.  This is a general problem for any model which includes significant deceleration, and thus we confidently eliminate {\it all} such models.
	
	\subsection{Simulations for Realistic Knots}
	
	The preceding sections offer a fairly idealized description of the radial distances of the knots from T Pyx as a function of time for various models.  But they ignore realities that inevitably complicate the picture.  The knots will presumably be ejected with a range of velocities, and the velocity vectors will have a range of angles to the line of sight.  Potentially variable deceleration amounts could lead to the intermingling of knots from different nova events within the visible shell.  These effects will complicate the simple picture displayed in Figure 6.  Our goal in this subsection is to simulate realistic knots and produce similar figures for comparison.
	
	The simulation of the knots is easy for the case without deceleration.  We simulate a number of knots (30 is reasonable to compare with the data) from two different eruptions.  For each simulated knot, we have randomly selected a velocity in some range (from $V_{min}$ to $V_{max}$) and a random direction on the sky.  For this knot, we then calculate the angle from the line of sight, the radial distances from T Pyx in the years 1994.1, 1995.8, and 2007.5, and the apparent angular separation of the knot from T Pyx as it appears on the sky.  The angular separations are converted to {\it HST} pixels for direct comparison with our images.  We also keep track of the position angles from T Pyx so as to draw simulated pictures of these knots on the sky.  From this, we can construct a plot of the motion between 1995.8 and 2007.5 as a function of the 2007.5 angular separation.  
	
	A sample of this is shown in Figure 8.  This simulation is for 30 knots ejected in 1866 with velocities of 500-715 km s$^{-1}$ and also for 30 knots ejected in 1967 with velocities of 1500-2000 km s$^{-1}$.  We have selected these velocity ranges to roughly reproduce the observed angular distance distribution for the observed knots.  (Nevertheless, even at 2000 km s$^{-1}$, the 1967 ejecta cannot get far enough away from T Pyx.)
	
	The ejecta from any given eruption always fall exactly on a straight line pointing back to the origin.  The slope of that line will immediately give the date of the nova eruption.  Any range of original velocities and angles to the line of sight will only shift the position along the line.  Different eruptions are clearly separated out as all the knots lie along lines of significantly different slope.  With this, we have a simple means of determining whether the knots come from multiple eruptions and have since intermingled together.  An examination of Figure 6 does not show evidence of multiple eruptions.
	
	The distribution of the knots along the line can give us statistical information about the distribution of velocities.  In Figure 8, we see an inner edge to the knots that corresponds to some fraction of $V_{min}$, while the outer edge corresponds to nearly the value of $V_{max}$.  For our assumptions of random ejection directions and velocities within a range, the distribution of points along the lines is roughly uniform.  In principle, the data in Figure 6 can be worked backwards into a velocity distribution.  Earlier in this paper, we have reported our Monte Carlo simulations, from which we deduce that the knots have velocities ranging from roughly 500-715 km s$^{-1}$.
	
	The effects of deceleration will be readily seen in this figure.  If all the knots decelerate by a consistent amount, they will still follow a straight line in the figure, with the indicated eruption date corresponding to the extrapolated date based on the velocity from 1995.8 to 2007.5.  All of the positions will be {\it below} the non-decelerated line in the figure.  The trouble then is that all models invoking significant deceleration of the knots are such that the degree of deceleration depends greatly on the highly variable mass and cross section of the knots.  A massive knot with much inertia will be slowed little and will be near the line, while a small mass knot will be greatly decelerated and fall far below the line.  A hallmark of any deceleration model, therefore, is that the distribution of points in the figure must be greatly spread out vertically.
	
	Two of the knots (K18 and K22) are moderately below the line defined by the other knots.  While it is possible that these two knots have undergone significant deceleration or were ejected in an earlier eruption (apparently in the first half of the 1800s), we consider the more likely scenario that the 1995 position of K22 is actually that of an unrelated, much fainter nearby knot, and that the 1995 position of K18 has been confused with the many knots in the area.
	
\section{Powering the Knots}
	
	A variety of explanations have been proposed to explain why the knots are shining as well as the source of their radiated energy.  The first possibility is that the knots are still ionized and glowing from their original eruption, presumably around 1866.  The mechanism for this is that they were ejected as ionized gas and are still recombining, which produces the usual emission lines.  The recombination time scale for hydrogen gas is $100,000/N_e$ years, where $N_e$ is the number density of the free electrons (Osterbrock 1989, p. 17).  Moderate densities allow the knots to keep shining after a century.
	
	The second possibility is that the knots are pre-existing and are powered by the light of some later eruption.  From the analysis in the previous section, it appears that the knots were ejected during an eruption around 1866 and moved out to their current position, where they could be made to glow by the light from the 1967 eruption.  In principle, the emission of light from the knots could be powered either by flash ionization (the ultraviolet light ionizes the gas which then emits lines as the atoms recombine) or by reflection off dust as in a reflection nebula.  (If dust is present in any significant amounts, then the dust would also radiate in the infrared.)  But the shell's spectrum is dominated by emission lines and not by a continuum, so the dust reflection idea can be rejected.  A case of flash ionization of a pre-existing shell was observed for nova V458 Vul (Wesson et al. 2009).  The light travel time through the shell is close to a quarter of a year, so all the knots would have had to turn on in 1967 at the latest.
	
	A third possibility is that the knots are pre-existing and are powered by a highly luminous and hot white dwarf.  Indeed, Knigge et al. (2000) make a case that T Pyx is a highly luminous supersoft X-ray source (but see Selvelli et al. 2008), and this is exactly what is needed to keep the knots shining.  The situation would be similar to that of a planetary nebula, in which the gas is being lit by the hot white dwarf at the center, and the recombination of the electrons gives off the observed emission lines.  This possibility has credence due to the analysis by Williams (1982) that the knots' emission lines are ``very similar" to those of planetary nebulae.  Presumably, the second and third possibilities can be distinguished because the temperature of the ionizing source would be greatly different in the two cases, and this would result in calculable and different ionization conditions in the knots and hence in different emission line strengths.  The supersoft source presumably provides continuous illumination, so the knots should have turned on long ago and stayed on.
	
	A fourth possibility is that the knots are powered by collisions between shells from different eruptions.  This possibility is well-supported by the quantitative analysis of Contini \& Prialnik (1997), which finds that the six observed emission lines from the knots are well matched by the emission expected from shocks in knots caused by collisions.  Their presumption was that the faster 1967 ejecta were running into the slower 1944 ejecta.  But this could just as easily be from the 1967 and 1944 ejecta running into the slower `1866' ejecta.  Within this model, each knot will light up when the later ejecta run into it, and the knot will then slowly fade as its electrons recombine.
	
	\subsection{Knots that Brighten}
	
	In our detailed comparison of knots between the 1995 and 2007 images, we see both a general and specific correspondence between knots in the two images.  But some of the knots have substantially brightened from 1995 to 2007.  The most obvious case is for K7 (see Table 1) which is one of the brightest in 2007 but is completely absent in 1995.  This knot underwent an event after 1995 that caused it to turn on.  Of the 30 knots visible in 2007 (Table 1), five are apparently invisible in 1995 while nine brighten by more than one magnitude from 1995 to 2007.  A similar result was already known from a comparison of the 1994 and 1995 {\it HST} images for three knots (Shara et al. 1997).  Bright knots are turning on at the rate of roughly one per year.  At this rate, the 30 (relatively bright) knots visible in 2007 should have all turned on over the past few decades.
	
	The observation that knots are turning on between 1994 and 2007 immediately allows us to determine the emission mechanism.  The first three mechanisms for powering the knots cannot allow for the turning on of the knots many decades after the last eruption.  In particular, knots powered by the original eruption will turn on at the time of the eruption and only fade thereafter, knots powered by a later eruption will turn on at the time of the later eruption and only fade thereafter, and knots powered by the supersoft source around the white dwarf will remain at a steady brightness.  In contrast, knots powered by collisions with later ejecta will suddenly light up over a wide range of times as the knots of various velocities finally collide to produce shocked gas.  Thus, as the only non-rejected possibility, we conclude that the knots are powered by shock heating caused by collisions between ejecta from different eruptions.
	
	In the previous section, we found that the knots were ejected at relatively low velocity by an eruption around 1866, and one of the later eruptions can provide the faster ejecta that powers the shocks.  To evaluate this, it is important to note that the shell was already visible with a similar shape in 1979 (Duerbeck \& Seitter 1979) and 1980 (Williams 1982).  The bulk of the material ejected in the 1967 eruption had velocity of close to 2000 km s$^{-1}$ and none had higher velocity (Catchpole 1969), while we measure the expansion velocity of the knots from the `1866' eruption to be around 600 km s$^{-1}$.  With these velocities, the collisions will only happen in the year 2005.  Thus, the simple scenario (1967 ejecta colliding with 1866 ejecta) does not work to make the shell bright in 1979.  If instead we use the 1944 eruption, the collisions happen in the year 1972, so the shell would be bright by the year 1979.  The 1920 eruption causes collisions around the year 1938.
	
	We must ascertain whether the relatively small collisional mass from the later eruptions can ionize the knots.  We assume that the shock velocity ($V_{shock}$ in units of km  s$^{-1}$) is the relative impact velocity (2000-600=1400 km s$^{-1}$). The downstream temperature, close to the shock-front, which is given by $T=1.5 \times 10^5(V_{shock}/100)^2$ K, is $2.9 \times 10^7$ K, leading to full ionization of the gas throughout the knots, depending on their density and geometrical thickness. 
	
	We must also take into account dispersion in the velocities.  The 1967 eruption does not show velocities greater than 2000 km s$^{-1}$, so the average turn-on times will not be before the year 2005.  We would need velocities from the 1866 knots as low as 200 km s$^{-1}$ to get collisions in 1979 from the 1967 eruption.  This is not a reasonable possibility, as then too many knots would be too close to T Pyx, which does not match the observation that the majority of the knots are at $\gtrsim100$ pixels as observed in 2007.  For a reasonable range of velocities (400-800 km s$^{-1}$) for the 1866 knots, we get collisions from 1989 to 2026.  Thus, the 1967 eruption cannot account for the shell being bright in 1979.  For the same range of velocities, the 1944 eruption will turn knots on from 1961 to 1988, and can account for the shell brightness when it was discovered.  For the same velocities, the 1920 ejecta collide with the 1866 knots from 1931 to 1948; those knots would have largely faded away by 1979.  Thus, the knots seen to be bright in 1979 were shining because of collisions with the 1944 ejecta, while the 1967 ejecta powered the knots that brightened from 1994 to 2007.
	
	\subsection{Knots that Fade}
	
	From 1995 to 2007, 5 of the 30 knots visible in 2007 faded by more than 0.5 mag.  For example, knots K21 and K15 (see Table 1; these are knots A and B of Shara et al. 1997) suffered their collisions between 1994.1 and 1995.8 and then faded by 1.59 and 1.21 mags respectively.  The simple interpretation is that this fading is caused by the decline in the electron number density due to electron recombination after a single shock episode some number of years in the past. This scenario requires that the recombination length downstream of the shock be comparable or smaller than the geometrical thickness of the knot, with this requirement being realized for typical estimated knot sizes. 
	
	Within this interpretation, we can relate the decline rate to the electron density.  The time scale $T_{rec}=100,000/N_e$ years (with electron density $N_e$ in units of cm$^{-3}$) is the e-folding time over which the electrons will recombine.  This standard equation actually has a temperature dependance, with the relevant temperature being relatively high close to the shock front leading to low recombination coefficients, and hence to a somewhat longer recombination time scale.  The knot brightness will be proportional to $N_eN_H$, where $N_H$ is the hydrogen density.  For a completely ionized gas, the knot brightness will be proportional to $N_e^2$, while this brightness will go down as the electrons recombine over time.  Over a time $T_{rec}$, $N_e$ will fall by a factor of 2.7, the flux from the knot will fall by a factor of 7.4, and the knot's magnitude will fade by 2.2 mag.  In general, for the simple case where a knot fades by $\Delta m$ over $\Delta Y$ years, the electron number density will be 
\begin{equation}
N_e = (\Delta m / 2.2 ~ {\rm mag})(100,000 ~ {\rm y}/ \Delta Y), 
\end{equation}
with units of cm$^{-3}$.  For knots K21 and K15 (with $\Delta m \approx 1.5$ mag over $\Delta Y = 11.7$ years), we have $N_e \approx 6000$ cm$^{-3}$.  It is encouraging that this density is close to that estimated by both Shara et al. (1997) and Contini \& Prialnik (1997).  For the electron densities required so as to give the observed knot luminosity, we expect a fading rate of order 1.5 mag in 11.7 years.

	This picture of fading will have various complications.  First, the densities within each knot undoubtedly vary substantially, so relatively sparse sections of the knot will decline slowly in brightness while relatively dense volumes will fade more quickly, with the result that the fading is only approximately linear for some intermediate density.  Second, the collection of visible knots will undoubtedly have a range of effective densities, so we will see a range of roughly linear decline rates.  Third, the ejecta from the later eruptions could be clumpy and will probably have some sort of velocity dispersion, so the shocks that excite any given knot will likely last for a long time, resulting in some complex competition between brightening and fading.  Indeed, this third complication is a reasonable explanation for why some knots (like knot C of Shara et al. 1997) display extended turn ons from 1994.1 to 1995.8, as well as the moderate brightening for many knots (see Table 1) from 1995.8 to 2007.5.  In all, the derived $N_e$ values can only be regarded as reasonable approximations.
	
	In summary, the knots are certainly powered by collisional shocks when gas clumps emitted in the `1866' event are hit by the ejecta from later eruptions.  Collisions with the 1944 material are largely what made the shell bright around the time of its discovery in 1979.  Since then, the collisions that are causing the individual knots to turn on have been with the material from the 1967 eruption.  After each collision, the ionized gas starts recombining on a time scale that produces a fading of roughly 1-2 mag per decade, corresponding to an electron density of $\sim 6000$ cm$^{-3}$.
	
\section{Shell Structure}
	
	The most prominent feature of the shell is the swarm of unresolved knots that are clustered in broad arcs at a distance of around 5" from T Pyx.  Other features have also been described.  Shara et al. (1989) report on the discovery of a faint outer halo visible to roughly 10" from T Pyx.  In addition, various papers have identified local maxima in radial profile plots as separate shells, including a 2" ring (Shara et al. 1989),  two maxima associated with the 1920 and 1902 eruptions (Seitter 1986), and nine shells (Shara et al. 1997).
	
	Our {\it HST} images show details of the shell structure.  To evaluate this, we have measured the flux in annular rings in two ways.  The first way, we add up all the flux within each annulus.  In the second way, we take the mode of all the pixels within the annulus to get the smoothed surface brightness without any flux from the knots.  For both methods, we subtract out the presumed constant background (as evaluated in the corners of the {\it HST} image, more than 600 pixels away from T Pyx) and the scattered light from T Pyx itself (as evaluated by scaling the flux around the Check star in similar annular rings).  For both methods, we report the surface brightness in units of 0.001 counts per pixel.  For calibration, the Check star (with V=15.81, R=15.33, and I=14.95) has a total of 83.4 counts inside a radius of 15 pixels.  For both methods, we measure the radial profiles with images made from both of our orbits in 2007, and the differences then provide a measure of the uncertainties arising from photon statistics and image artifacts.  The result is that the one-sigma uncertainty on the average surface brightness is 0.18-0.19 for radial distances $<150$ pixels and 0.07-0.10 outside that (in units of 0.001 counts per pixel).  Our two radial profiles are presented in Figure 3.
	
	The most prominent feature in our radial profiles is  the relatively bright structure from roughly 25-150 pixels (1.1-6.7"), with a distinct peak near 110 pixels (5.0").  This is caused by the bright, slowly-expanding knots of the `1866' ejecta.  The mode profile parallels the total flux profile; this is likely caused by the myriad of unresolved knots.  That is, we know from the much deeper 1995 {\it HST} images that there are over 2000 knots (Shara et al. 1997), most of which must be unresolved in our 2007 image.  The peak around 110 pixels is exactly as expected for a spherical shell of knots sent out with a similar velocity.  If all the knots have the same velocity, then the cusp will be fairly sharp.  The fall-off above the peak is determined largely by the velocity dispersion, such that the observed cutoff implies a 500-715 km s$^{-1}$ range in velocities.  Again, the uncertainty in this range is roughly 50 km s$^{-1}$.  We see structure in our radial profile with peaks at 35, 55, 75, 105, 185, and 235 pixels.  However, given the uncertainties in the measures and the random shot noise in knot positions, we do not think that any of the peaks (other than the high peak around 110 pixels) are significant or represent any shell.  In Monte Carlo simulations of radial profiles of knots ejected in one shell, we often see random apparent peaks that shift around from realization to realization.  We also do not see any correlation with prior profile peaks, so we conclude that the earlier identified peaks do not have any separate physical identity.  
	
	At radial distances greater than 180 pixels (8"), we see a significant flux above the background level.  This flux falls off inversely with distance from T Pyx.  Roughly, the surface brightness at 360 pixels (16") is half that of the surface brightness at 180 pixels.  However, we see substantial variations around any idealized power law decline.  For the surface brightness falling inversely with radius, this means that the total brightness in each annulus should be constant with radius.  For a presumed uniform velocity, this would imply that the mass ejection was uniform in time, or at least that the velocity spreads in successive eruptions have smoothed the outflow to a fairly uniform expansion.
	
	The obvious explanation for the cause of this outer halo is that the fast-moving gas from the many RN events that has reached large distances from T Pyx by passing the slow knots from the `1866' nova event.  With the slow moving `1866' ejecta having broken up into knots of relatively small surface area, most of the later ejecta will pass by.  The interaction of all these shells (with their knots and spatial inhomogeneities) will result in collisions, complex shocks, instabilities, and fragmentation (see for example Zanstra 1955; Graham \& Zhang 2000).  This paper is not the place to perform such realistic calculations, but we expect that the many RN shells will overlap and jumble together (especially as viewed in projection on the sky) to yield a radial profile much as observed in Figure 9.  For a transverse velocity of 2000 km s$^{-1}$, the gas will expand by a radial distance of 1.2" (25 pixels) in a decade, or it will move 4.8" (100 pixels) in nearly four decades (all for an adopted distance of 3,500 pc).  Our 2007 {\it HST} image was taken four decades after the 1967 eruption, and the gas is lighting up knots $\sim$100 pixels away from T Pyx.  The 1944, 1920, 1902, and 1890 eruptions should have gas out to 7.6", 10.4", 12.6", and 14.0" (157, 217, 262, and 292 pixels), respectively.  It is tempting to identify various small (i.e., insignificant) bumps in our radial profiles with these particular shells, and this might even be correct, but we feel that the confidence level in any such identification is too low to be useful.
	
	Thus far, only one attempt has been made to explain the most unique feature of the shell, the large number of distinct knots.  Garcia-Segura et al. (2004) modeled the hydrodynamics of repeated RN shells with velocity dispersions running into each other.  They found turbulence which does not look like the observed knots, and their calculations are not relevant for the case where the knots are from an ordinary nova event long ago with high ejecta mass.  With our conclusion that the knots are from an old shell being blasted by fast winds from later eruptions, we have an obvious explanation since the knots are shaped much like the so-called cometary knots in planetary nebulae.  The fast winds break up the pre-existing shell into knots due to the Rayleigh-Taylor instability.  This mechanism has long been known in the context of forming knots in expanding shells (e.g., Zanstra 1955) and has been intensively studied in general (e.g., Vishniac 1994), in the context of Wolf-Rayet stars (e.g., Garcia-Segura \& Mac Low 1995), and in the context of planetary nebulae (e.g., Capriotti \& Kendall 2006).  A particularly evocative picture of this case is the cometary knots in the nearby Helix Nebula, wherein a dense and slow-moving shell (from the planetary nebula ejection) is broken up into a myriad of knots by means of a Rayleigh-Taylor instability produced by a later fast-moving wind (O'Dell \& Handron 1996; Capriotti \& Kendall 2006).  The knots in the T Pyx shell are the same as those in the Helix Nebula except that we do not have the ability to resolve the knots, nor the sensitivity to see the cometary tails.  In all, the Rayleigh-Taylor instability and knot formation are expected when a thick and slow shell has a fast wind roaring through it.  We can be confident that the `1866' shell has been broken up and compressed into knots by the ejecta from later eruptions.
	
\section{Masses}
		
	\subsection{Shell Mass from the H$\alpha$ Flux}
	
	The flux in emission lines (like H$\alpha$) from all the knots should be proportional to the mass of the shell, or at least the ionized fraction of that mass.  The standard method equates the shell mass to a product involving the total line flux, various atomic constants, the square of the distance, and average densities of hydrogen and free electrons (Gallagher \& Starrfield 1978).  Unfortunately, this method is highly model-dependent, requiring assumptions on the poorly known distance, the filling factor, the composition, and the presence of a steady state.  Typical uncertainties in these assumptions lead to order of magnitude errors, so it is not surprising that published values for individual nova shell masses differ by over two orders of magnitude  (Schwarz 2002; Shaviv 2002; Vanlandingham et al. 2002: Gallagher \& Starrfield 1978).  Theoretical models have similar scatter.  We can only agree with the Gallagher \& Starrfield review that ``The masses of material expelled by active novae are poorly known."
	
	Nevertheless, an order of magnitude estimate is still useful.  To this end, Shara et al. (1997) performed a detailed analysis and concluded that the shell has a mass of $1.3 \times 10^{-6}$ M$_{\odot}$.  The explicit assumptions are that the distance is 1,500 pc, and $N_H=N_e=10^3$ cm$^{-3}$ where $N_H$ is the hydrogen number density.  We will update this calculation based on new information from this paper.
	
	One update is simply to use a possibly better distance, 3,500 pc instead of 1,500 pc.  With this, the derived shell mass increases to $7 \times 10^{-6}~ d_{3500}^2$ M$_{\odot}$.
	
	The second update comes from the realization that at any given time, not all knots are shining with peak brightness, so only a fraction of the gas is counted.  The knots that have turned on after 1995 were not counted, so the real shell mass must be somewhat higher than could have been known.  We can correct for this now that we have some idea of the turn-on rate of the knots.  Also, once a knot is shocked (presumably to full ionization) the recombination will reduce $N_e$ and hence the flux, resulting in an underestimate of the total mass in the knot.  That is, the H$\alpha$ flux at any one time only represents the ionized mass.  If a knot was shocked to full ionization a decade before 1995, then its H$\alpha$ brightness has fallen by a factor of $\sim$4 and we must correct the prior estimate by this factor for that knot.
	
	This correction factor will necessarily be complex as it incorporates the history of collisions and the ionization fraction of all the knots.  Nevertheless, we can estimate this factor.  To do this, we have constructed many Monte Carlo simulations in which we build up a shell from knots whose properties follow reasonable distributions and then follow each knot through its collision and brightness history.  For each simulation, we compare the flux from the case where all the knots are completely ionized to the flux from the case where the knots' brightnesses in 1995 are all summed.  This will produce a correction factor to the shell mass based on the flux in 1995 alone.
	
	The simulation follows 500 knots emitted in 1866 (with a range of velocities and masses) and then presumes that they are completely ionized when they are struck by ejecta from the 1944 and 1967 eruptions.  The knot velocities were taken randomly from a uniform distribution between 500 and 715 km s$^{-1}$.  The knot masses were selected randomly from a power-law distribution chosen so as to reproduce the observed knot luminosity function (Shara et al. 1997).  We also randomly chose a direction in space for each knot, which allows us to calculate radial distributions for the knots for comparison with the observed radial distribution in 1995 (Shara et al. 1997).  The 1944 and 1967 ejecta velocities in the direction of each knot were taken randomly from a uniform distribution between 1500 and 2000 km s$^{-1}$.  For each knot, we calculated the year of impact for each eruption.  The brightness of each knot at the time of impact was assumed to come from full ionization, with recombination resulting in the fading of each knot at a rate of 0.13 mag per year. This scenario matches the observations for the knot radial distribution, the motion, the frequency of turn-on, and the luminosity function in 1995.  There are many complexities which are not accounted for in this simulation, including the effects of potential partial knot ionization, non-instantaneous collisions, variable density throughout each knot and from knot to knot, and the presence of helium and metals in the gas.  Nevertheless, the model correction factor is approximately correct and certainly better than no correction at all.  With these simulations, we find that the correction factor is $4.4\pm0.3$, where the uncertainty represents the variation between individual Monte Carlo realizations.  This factor changes from roughly 1.35 to 17 as we vary the input assumptions over plausible ranges.  With our best correction factor of 4.4, the shell mass becomes $3.1 \times 10^{-5} ~d_{3500}^2$ M$_{\odot}$.  Given the size of the uncertainties, perhaps this is better expressed as $\sim 10^{-4.5}$ M$_{\odot}$ for the total mass in the knots from the 1866 eruption.
	
	Despite the order of magnitude nature of our estimate, this result immediately tells us that the 1866 eruption is greatly different from the observed recurrent nova events from 1890 to 1967, because the ejected mass would be much larger than is possible for the later eruptions.  With an accretion rate of $\sim 10^{-7}$ M$_{\odot}$ yr$^{-1}$ and a recurrence time of 12-24 years, the available mass is more than an order of magnitude too small to account for the observed mass.  Again we find that the properties of the 1866 eruption are greatly different from those of the later RN events.
	
	\subsection{Mass of the Recurrent Nova Shells}
	
	Selvelli et al. (2008) made several estimates of the mass ejected during the 1967 eruption and concluded that $M_{ejecta} \sim 10^{-4} - 10^{-5}$ M$_{\odot}$.  Unfortunately, these estimates have very large uncertainties and questionable assumptions.  One of their estimations of $M_{ejecta}$ comes from a correlation with $t_2$ by Della Valle et al. (2002).  But the existence of this correlation is {\it not} significant.  The correlation has a scatter of two orders of magnitude in $M_{ejecta}$, and was calibrated with ordinary novae whose white dwarf mass is generally quite low, whereas a high-mass RN white dwarf could have a greatly different calibration.  In all, no confidence can be attached to the estimate of the T Pyx shell mass based on the $t_2$ correlation.  Another of their estimates has the shell mass equaling the hydrogen density times the mass of a hydrogen atom times the volume of the shell, with a resulting estimate of $1.5 \times 10^{-4}$ M$_{\odot}$.  Such an estimate requires the average electron density, which cannot be accurately known at any time.  Also, the estimated shell mass is highly sensitive (as a cubic power) to the adopted shell size.  For this, Selvelli et al. simply took seemingly arbitrary choices of the expansion velocity times the $t_2$ value.  Additionally, this does not account for any recession of the photosphere, which is relevant as they are trying to model the optically thick portion alone.  In all, with the arbitrary choices and questionable assumptions, the derived shell mass has uncertainties of many orders of magnitude.  The other estimations of $M_{ejecta}$ have comparable problems, so we conclude that these estimates are not useful.
	
	Selvelli et al. (2008) were aware of these and other problems with their high estimates of $M_{ejecta}$.  They highlighted the discrepancy between their $M_{ejecta}$ and the trigger mass as well as with the accreted mass.  They also point out the dilemma that the ejecta abundances are near solar, which requires that no material is dredged up from the white dwarf.  In all, their arguments and analysis all point against their high $M_{ejecta}$ estimates.
	
	We know that the ejecta mass for the RN eruptions is low (of order $2 \times 10^{-6}$ M$_{\odot}$) for many good reasons.  First, the spectra during outburst show near solar abundances, which demonstrates that the dredge-up of material is negligible.  This implies that the ejected material must be limited to the accreted material, which is roughly 20 years times $10^{-7}$ M$_{\odot}$ yr$^{-1}$.  Second the ejected material should not be much greater than the trigger mass ($2 \times 10^{-6}$ M$_{\odot}$) because there is no time for the accreted material to diffuse below the old surface of the white dwarf.  Third, we do not see knots associated with any of the RN eruptions, and this implies that their ejecta are much less massive than the $\sim 10^{-4.5}$ M$_{\odot}$ in the ejecta from the `1866' event.
	
	The small $M_{ejecta}$ for the RN eruptions is an important issue for determining whether the white dwarf is gaining or losing mass over each outburst cycle.  Unfortunately, we do not have any direct measure of the ejecta mass, and the estimates from the accreted mass and the trigger mass are both only approximate.  Indeed, the use of the accreted mass would be circular in any question of whether the white dwarf is gaining or losing mass, and the use of the trigger mass would be circular in any question of testing models, as then the new model would simply be tested against an old model instead of against observations.  We are left with the unsatisfying conclusion that the mass ejected by the RN events is small ($\sim 2 \times 10^{-6}$ M$_{\odot}$) and poorly known.
	
	\subsection{Mass of the White Dwarf}
	
	Knowing the mass of the WD is critical for understanding the details of the eruption, the long-term evolution, and whether T Pyx can become a Type Ia supernova.  Unfortunately, $M_{WD}$ is not accurately known.  Observationally, the radial velocity is small and there is no useful constraint on the WD mass (Uthas 2009).  There is a strong theoretical imperative to have the mass in the range 1.2-1.4 M$_{\odot}$ so that the recurrence time scale can be sufficiently short.  Kato (1990) presented theoretical models that place the mass in the range 1.30-1.37 M$_{\odot}$.  Selvelli et al. (2008) examined a range of input (theoretical and observational) to reach the conclusion that $1.25<M_{WD}<1.4$ M$_{\odot}$, with a preference for the upper end of the range, perhaps $\sim$1.36 M$_{\odot}$.  They made detailed use of the latest theoretical models on nova triggering and eruptions as given by Yaron et al. (2005), plus some privately communicated models from Yaron et al.  We will repeat this analysis for comparing observations with the Yaron et al. models, including our new information about the `1866' nova.
	
	The first case we will examine is that of the classical nova eruption around the year 1866, for which the accretion was driven by gravitational radiation alone ($\dot{M}\approx 4 \times 10^{-11}$ M$_{\odot}$ yr$^{-1}$).  The measured quantities that we are trying to match are the total ejecta mass of $\sim10^{-4.5}$ M$_{\odot}$ and the maximum expansion velocity of the ejecta of 715 km s$^{-1}$.  Yaron et al. (2005) show that we cannot get the high observed mass in the knots for either high $M_{WD}$ or relatively hot WDs.  Indeed, we get a match only for the $M_{WD}=1.25$ M$_{\odot}$ models with the WD core temperature at ten million degrees.  To match the observed maximum expansion velocity, we have to go to $M_{WD}$ smaller than 1.25 M$_{\odot}$ for any core temperature.  So what we know about the `1866' eruption is strongly pointing to the lower end of the range (1.2-1.25 M$_{\odot}$).
	
	The second case is that of the RN eruptions, for which we will take the conditions around the middle of the 1900s when the accretion rate was within a factor of three of $3 \times 10^{-8}$ M$_{\odot}$ yr$^{-1}$ (see below).  The measured quantities we will try to match with the model of Yaron et al. (2005) are the recurrence time scale of 23 years, the maximum expansion velocity of 2000 km s$^{-1}$, the amplitude of 8.3 mag, and $t_3$ of 62 days.  The model values are only weakly dependent on the WD core temperature.  To match the observed recurrence time scale, the WD mass can be from 1.25 M$_{\odot}$ (for $\dot{M}= 10^{-7}$ M$_{\odot}$ yr$^{-1}$) to 1.40 M$_{\odot}$ (for $\dot{M}= 10^{-8}$ M$_{\odot}$ yr$^{-1}$).  To match the observed maximum velocity, the WD mass can be from 1.30 M$_{\odot}$ (for $\dot{M}= 10^{-8}$ M$_{\odot}$ yr$^{-1}$) to 1.40 M$_{\odot}$ (for $\dot{M}= 3 \times10^{-8}$ M$_{\odot}$ yr$^{-1}$).  To match the observed amplitude, the WD mass can be from 1.25 M$_{\odot}$ (for $\dot{M}= 3\times10^{-8}$ M$_{\odot}$ yr$^{-1}$) to 1.40 M$_{\odot}$ (for $\dot{M}= 2\times10^{-8}$ M$_{\odot}$ yr$^{-1}$).  To match the observed $t_3$, the WD mass must be around 1.25 M$_{\odot}$.  
	
	For these two cases, the preferred loci in the $M_{WD}-\dot{M}$ parameter space do {\it not} cross.  This leaves us in a quandry as to how to optimize the model to the observations.  A plausible choice would be to pick $M_{WD}$=1.25 M$_{\odot}$ and $\dot{M}= 10^{-7}$ M$_{\odot}$ yr$^{-1}$, but then the predicted maximum velocity of ejection for the RN events is wrong and the amplitude a bit small.  Another plausible choice would be to pick $M_{WD}$=1.30 M$_{\odot}$ and $\dot{M}= 7\times10^{-8}$ M$_{\odot}$ yr$^{-1}$, but then the predicted maximum velocity of ejection is wrong for both the RN and `1866' events, plus the RN amplitude and the `1866' ejecta mass are too small.  Our conclusions are in agreement with those of Selvelli et al. (2008), except that the `1866' event is pointing to the lower part of their allowed range for the WD mass.  In all, we take the theoretical models of Yaron et al. (2005) to point to $M_{WD}$=1.25-1.30 M$_{\odot}$ and $\dot{M}$= 7-10$\times10^{-8}$ M$_{\odot}$ yr$^{-1}$.
	
	\subsection{Trigger Mass}
	
	Detailed models can calculate the trigger mass, $M_{trig}$, that must accrete onto a white dwarf to ignite the thermonuclear runaway (Nomoto 1982; Townsley \& Bildsten 2005; Yaron et al. 2005; Nomoto et al. 2007; Shen \& Bildsten 2007).  These sources differ by up to a factor of 20 for the relevant cases, but we will generally be guided by the results of Yaron et al. (2005) which has the most complete physics for the nova calculations.  The trigger mass will be a function of the mass of the WD and the accretion rate.  It should be comparable to the ejected mass, with the ejected mass equaling the trigger mass to within a factor of 0.6-1.2 for the situations of interest (Yaron et al. 2005).  Though not exact, we can still get an idea of what is going on.
	
	One situation of interest is for the RN phase during the middle 1900's.  During this time, the accretion rate is likely $\sim10^{-7}$ M$_{\odot}$ yr$^{-1}$ and the WD mass is around 1.25 M$_{\odot}$.  With interpolation and extrapolation as needed, the calculated $\log(M_{trig})$ is -5.7, -5.5, $\sim$-6, -6.5, and -5.3 for the references Yaron et al. (2005), Nomoto (1982), Townsley \& Bildsten (2005), Nomoto et al. (2007), and Shen \& Bildsten (2007), respectively.  We adopt the value from Yaron et al. (2005) which is in the middle of the range, with a trigger mass of $2\times10^{-6}$ M$_{\odot}$.  As a consistency check, we recover a recurrence time scale of 20 years from $M_{trig}/\dot{M}$.  From the theoretical models of Yaron et al. (2005), the ejecta mass will be about 10\% less than the trigger mass.  So we now have an approximate value for the mass ejected during the various RN events of the 1900s.  This ejected mass is much smaller than our estimated mass in the knots, with the confirmatory implication that the RNe could not have ejected the currently visible knots.  Additionally, the RN ejecta have so much less mass than the knots that they should not be visible on their own.  A third conclusion is that there will be no dredge-up of material during the RN eruptions.
	
	The other situation of interest is for the ordinary nova event around 1866.  Here, we know that $M_{ejecta}\sim10^{-4.5}$ M$_{\odot}$, and the trigger mass is approximately the same.  The only way to get such a large trigger mass on the WD is to have a very low accretion rate.  This demonstrates that the `1866' event is not like the RN events.  To get a trigger mass of $10^{-4.5}$ M$_{\odot}$, the models of Nomoto (1982), Townsley \& Bildsten (2005), and Shen \& Bildsten (2007) suggest that the accretion rate is near $4 \times 10^{-11}$ M$_{\odot}$ yr$^{-1}$.  The model of Yaron et al. (2005) only produces such a high $M_{trig}$ for WDs in the lower part of the RN range, for the cooler WDs, and for lower accretion rates.  The model of Nomoto et al. (2007) produces such high $M_{trig}$ only for accretion rates below $\sim 10^{-11}$ M$_{\odot}$ yr$^{-1}$.  In all cases, it is clear that before the `1866' eruption, T Pyx had a very low accretion rate, roughly $4 \times 10^{-11}$ M$_{\odot}$ yr$^{-1}$ or lower.
	
	In ordinary circumstances, CVs cannot have accretion rates below the minimum provided by the angular momentum loss from gravitational radiation.  For T Pyx, this minimal value is $\dot{M}= 4 \times 10^{-11}$ M$_{\odot}$ yr$^{-1}$ (Knigge et al. 2000).  This limit matches well with the estimated accretion rate for the time interval before the `1866' ordinary nova eruption.    This concordance makes for a good conclusion that this is the real rate, and that T Pyx was an ordinary CV before the `1866' nova.  The required time interval for T Pyx to accumulate a trigger mass of $\sim10^{-4.5}$ M$_{\odot}$ while accreting at a rate of $4 \times 10^{-11}$ M$_{\odot}$ yr$^{-1}$ is around 750,000 years.  In all, for the three-quarters of a million years before 1866, T Pyx was an ordinary CV slowly accumulating material on the surface of its WD.
	
	The big eruption in $1866\pm5$ could not have been identical to the RN eruption observed in 1890.  Likely, T Pyx suffered one or two undetected RN outbursts between 1866 and 1890.  The observed very high accretion rate from before the 1890 eruption implies a fast recurrence time of perhaps a few years to a dozen years, but not 24 years.  Based on the 12-year interval from 1890 to 1902, it seems likely that the 24-year interval from 1866 to 1890 contained one eruption around the year 1878.  With a higher accretion rate, it is possible to have two RN eruptions between 1866 and 1890.
	
\section{The Secular Fading of T Pyx}
	
	We have direct observations of the quiescent B-band magnitude of T Pyx from {\it before} the 1890 eruption up until 2009.  Schaefer (2005) has collected the B-band magnitudes from the Harvard plates, literature (primarily from Landolt 1977; Schaefer et al. 1992 and Patterson et al. 1998), and his own unpublished CCD images from 1992 to 2004.  We have been continuing observations up to the time of this writing.  The most important result from this light curve is that T Pyx has been declining from B=13.8 mag to B=15.7 mag from 1890 to 2009.  This is a drop of 1.9 mag, which is a factor of 5.7 in B-band flux.
	
	This drop in brightness is unprecedented, unique, and highly significant.  Neither the WD nor the low-mass companion can provide any noticeable amount of light in the B-band, so the brightness must come entirely from the accretion disk, and variations in this brightness can only arise from changes in the accretion rate.  Schaefer (2005) has made detailed calculations for alpha-disk models of T Pyx which show that the accretion rate varies as  
\begin{equation}
\dot{M} \propto F^{2.0}, 
\end{equation}
where F is the B-band flux.  We have just shown that T Pyx dropped by a factor of 5.7 in flux, and hence the accretion rate has dropped by a factor of 33.  That is, the accretion rate in 2009 is only 3\% of that in early 1890.  This tremendous secular change must have an effect on the evolution of T Pyx.
	
\section{The Next Eruption of T Pyx}
	
	In the late 1980s, based on the last eruption being in 1967 and the average recurrence time scale of roughly 22 years, the common expectation was that T Pyx would go off any year.  But as the years stretched on with no eruption, hope faded, to be replaced by frustration and perplexity.  One important realization is that the intense monitoring of T Pyx combined with its long eruption duration means that no eruption could have slipped through the solar gap, which is to say that any T Pyx eruption {\it would} have been discovered had it occurred.
	
	Schaefer (2005) realized that T Pyx has been suffering a long decline in brightness, and hence its accretion rate has been largely turning off.  With a drop of 0.7 mag in brightness ($2\times$ in flux and $4\times$ in $\dot{M}$) across the 1967 eruption, we have an easy explanation for why the next eruption of T Pyx has been long delayed.  The long delay is due to the accretion rate falling by a factor of four (compared the to previous eruption cycle), so it must take four times longer for the trigger mass to accumulate.  Schaefer (2005) gave a more detailed calculation and concluded that the next eruption would be in the year $2052\pm3$.  This prediction is based on the assumption that T Pyx remains at its 2005 brightness level, $B=15.5$ mag.
	
	Selvelli et al. (2008) revisited this question.  They used the same idea, trying to estimate how long it would take the system to accrete the trigger mass.  They concluded that the next eruption would be in the year 2025, but stated no error bars.  Unfortunately, they tried to evaluate the quantities in an absolute sense, but even their extremely small quoted error bars for distance and WD mass lead to an uncertainty in their estimate of a factor of three.  As such, their one-sigma range is actually 2000-2067.  (The Schaefer 2005 result avoids these problems by combining all the unknowns into one factor that can be empirically calibrated from the prior eruptions, and as such is completely independent of any assumptions about the distance or WD mass.  With this, the real error bar will be greatly smaller than what is possible for the method of Selvelli et al.)  This is fully consistent with the Schaefer (2005) prediction of $2052\pm3$.  Selvelli et al. (2008) also present an analysis wherein they only scale from the 1944-1967 interval.  (This is a restricted version of the method from Schaefer 2005.)  They note the change in average brightness from 1944-1967 to after 1976 by a factor of nearly two, claim that $\dot{M}$ changed by a factor of two, add 50\% to account for the change in the trigger mass, conclude that the next inter-eruption interval will be 60 years, and add this to 1967 to get an eruption in the year 2025.  This analysis has a large error in taking the accretion rate to be proportional to the B-band flux.  They presume without justification that $\dot{M} \propto F$, rather than the result in equation 2.  But accretion disks are such that halving the observed flux corresponds to a quartering of the accretion rate.  Their estimated inter-eruption time should be doubled, putting the next eruption in the year $1967 + 4 \times 1.5 \times (1967-1944) = 2105$.
	
	Selvelli et al. (2008) make the good point that the trigger mass will significantly change due to the falling accretion rate, which we independently realized as well.  As such, the original prediction in Schaefer (2005) should be modified to account for this.  From Nomoto (1982), a change in $\dot{M}$ by a factor of four (with a white dwarf mass of 1.30 M$_{\odot}$) changes the trigger mass by a factor of about two.  Thus, the inter-eruption time interval for T Pyx should increase from $2052-1967=85$ years to roughly 170 years.  With this approximate correction, the estimated date of the next eruption is $1967+170=2137$.
	
	These predictions are based on the assumption that the accretion rate stays at the 1976-2004 level.  But T Pyx has started another fading episode, and has now declined to 15.71 (based on seven nights from 2009 March 3 to 2009 April 26).  Thus, the flux has fallen by 20\% and the accretion is now at 60\% of its 2004 level.  This will delay the next eruption yet more.  Formally, if T Pyx stays at the 2009 level, then the time from 2004 to the next eruption will be extended by a factor of 5/3, giving the next eruption in the year 2225.  Since the trigger mass has increased yet more due to this further decline, the next eruption must be substantially after the year 2225.  The inter-eruption time interval will then be $>259$ years, and we would question whether T Pyx can be called an RN anymore.
	
\section{The Cause of the High Accretion Rate}
	
	One of several key mysteries of T Pyx is the cause of its very high accretion rate.  As a cataclysmic variable {\it below} the period gap, T Pyx should have its accretion driven solely by the loss of angular momentum from gravitational radiation, and this should result in $\dot{M}=4 \times 10^{-11}$ M$_{\odot}$ yr$^{-1}$.  But T Pyx has an accretion rate that is currently higher by a factor of around a thousand.  There must be some additional mechanism driving the fast transfer of matter.
	
	We can conceive of a variety of explanations.  For example, we could postulate that the secondary star is just starting to evolve off the main sequence, with the expansion of the star powering the high accretion.  This is quickly refuted with the realization that the secondary star must have such a low mass that it will never leave the main sequence, even in many Hubble times.  Another idea might be that the orbital period is not the strongly-confirmed photometric period, but is instead longer.  For example, Selvelli et al. (2008) recognized a low probability that the true orbital period is the spectroscopic period of 3.439 hours reported by Vogt et al. (1990), but even with this period, the observed $\dot{M}$ is much too large.  In any case, the photometric period has been confirmed spectroscopically (Uthas 2009).
	
	With the need to explain the high accretion rate, the papers by Kovetz et al. (1988) and Knigge et al. (2000) offer a good solution.  Kovetz et al. (1988) showed that irradiated companions of nova WDs expand in size enough to drive mass transfer that is enhanced by two orders of magnitude over a century.  Knigge et al. (2000) proposed that T Pyx is a high-luminosity supersoft source: nuclear burning on the surface of the WD irradiates the secondary star and powers a wind with a high accretion rate.  This accretion would be high enough to sustain the luminosity of the supersoft source and keep a presumably steady state.  Their scenario is presented in greater detail for a general case by van Teesling \& King (1998).  The original cause of the supersoft case is unexplained, and the trigger event is described as an ``accident."  Knigge et al. (2000) are more explicit and discuss the possibility of the triggering event being residual nuclear burning near the end of a nova eruption on a high-mass WD with a low-mass companion.
	
	The self-sustaining, irradiation-induced accretion possibility requires fairly particular conditions (Frank et al. 2002, pp. 77-78).  For example, a high X-ray luminosity is not efficient at driving a wind, whereas a supersoft source is particularly effective.  Apparently the required conditions are rare.  But this need not worry us for T Pyx, as it is a unique and exceptional source, so we already know that we are dealing with a rare mechanism.
	
	Selvelli et al. (2008) give a detailed analysis based on {\it IUE} and {\it XMM-Newton} spectra that raises critical problems with the supersoft scenario.  (Historically, the same authors used similar arguments centered in {\it IUE} spectra to refute the idea that T CrB eruptions were caused by instabilities in the accretion disk, see Selvelli et al. 1992.)  They point out that the {\it IUE} spectral shape is that of an accretion disk and not of a supersoft source, that the ultraviolet+optical+infrared luminosity is  lower than given in Patterson et al. (1998), and that the emission lines are not those of high excitation.  In the year 2006, their {\it XMM-Newton} X-ray spectrum rules out a significant flux from a supersoft source.  With this, we have a strong case that T Pyx is not curently a supersoft source.
	
	It might be possible to reconcile the Selvelli et al. (2008) observations with the supersoft hypothesis, based on the stark secular decrease in the accretion rate.  That is, T Pyx might have been a bright supersoft source in the years soon after `1866', but the fall off in accretion means that it is no longer such a source.  The accretion rate has dropped to $\sim3\%$ of its value in 1890, so its current $\dot{M}$ should be too low to produce a detectable supersoft source.  So the Knigge et al. (2000) model is correct at explaining why the accretion rate is so high, while the Selvelli et al. (2008) observations are also correct in that T Pyx is not {\it now} a supersoft source.  The Selvelli et al. (2008) accretion rate is close to $10^{-8}$ M$_{\odot}$ yr$^{-1}$, and this is consistent with the lack of a detection in soft X-rays.  This also means that the accretion rate more than a century ago was roughly $3\times10^{-7}$ M$_{\odot}$ yr$^{-1}$, and this is just at the upper limit for the absence of steady burning on a high-mass WD.  The general reconciliation is that T Pyx was a bright supersoft source starting around 1866, but has been fading away as the accretion rate falls off so that it is not currently detectable in soft X-rays.  This idea requires that there must be some sort of time constant of decades for the response of the accretion rate to the stimulus of the irradiation.  With this reconciliation, the only modification to the original idea of Knigge et al. (2000) is that the accretion rate is not perfectly self-sustaining, but rather the feedback is imperfect and the accretion rate is slowly falling. 
	
	The only currently-acceptable explanation for the high accretion rate is the irradiation-induced accretion idea.  This mechanism is not completely worked out, but it is reasonable, and there is no alternative in sight.  We conclude that this is the likely cause for the high accretion rate in T Pyx.  The supersoft source and the induced accretion would have all started during the `1866' nova event, with this being responsible for the fast fall of matter onto the white dwarf and the subsequent recurrent nova eruptions.  The secular decline of quiescent brightness proves that the induced accretion is not perfectly self-sustaining.

\section{The Evolution of T Pyx}
	
	The two new key facts about T Pyx are the nature of what we are calling the `1866' event and the century-long decline in accretion since then.  This leads to a confident picture of what the system is like during its phase of nova eruptions (which we label as the `RN state').  The time before the `1866' eruption has T Pyx in a completely different state (which we label as the `ordinary CV state'), while the stopping of accretion implies that T Pyx will go into yet a third state (which we will label as the `hibernation state', cf. Shara et al. 1989 and the hibernation scenario of cataclysmic variables).  So we see the long term evolution of T Pyx as passing through at least three states in succession, from the ordinary CV state to the RN state to the hibernation state.  Transition from the deeply hibernating          
state to the CV state will involve increasing rates of mass transfer, possibly bringing on U Gem and Z Cam dwarf nova behavior, too (Shara et al. 2007).  Indeed, the hibernation state must inevitably change to the ordinary CV state, and so we have a cycle.
	
	\subsection{The Recurrent Nova State}
	
	We have shown that there was an ordinary nova eruption around the year 1866.  A shell of approximately $10^{-4.5}$ M$_{\odot}$ was ejected.  The knots in that shell have expansion velocities in the range 500-715 km s$^{-1}$ and approximately solar metallicity, so they consist mostly of matter that had been transferred from the companion star, not dredged-up WD material.  After this eruption, T Pyx entered a state with a very high accretion rate, likely caused by nuclear burning on the surface of the WD.  In ordinary classical novae, the nuclear burning of the supersoft phases turns off relatively quickly, but it was sustained in T Pyx because of the high-mass WD and the proximity of the secondary star.  This is a rare situation because only a small fraction of nova systems have high-mass WDs in addition to very short orbital periods.
		
	The observed high accretion rate determines the properties of T Pyx during the RN state, including the relative quiescent brightness.  Since material is accreting quickly, T Pyx has a low trigger mass and fast recurrence time scale.  The secular brightness decline, however, can only be caused by a corresponding decline in the accretion rate, which is currently at 3\% of its rate before the 1890 eruption.  Because of this, T Pyx no longer has a significant supersoft luminosity (Selvelli et al. 2008) and has a much longer recurrence time scale.  The supersoft source was not strong enough to drive a self-sustaining accretion flow, so the decline in magnitude and accretion rate will continue, and T Pyx will not erupt again in the foreseeable future.
	
	Over the last 120 years, $\dot{M}$ has declined by a factor of 33.  In the simple extrapolation for a decline at this rate, it will take another two centuries for the accretion to fall from its current level down to that provided by the usual rate forced by gravitational radiation.  But the real situation that must be considered when attempting to determine the length of the RN state is how long the now-puffed-up companion will remain with a significant atmospheric density in contact with the Roche lobe.  This would account for the latency of the heat deposited earlier in the companion (which causes the expansion of the star) and the scale height of its atmosphere.  We do not have an answer to this question, but we suspect that the time will be less than that given by the naive extrapolation.  We adopt a total time in the RN state ($\Delta T$) of two centuries, although none of our results depend significantly on this assumption.

	During the RN state, the fast rate of mass transfer will cause changes in the total mass of both stars in the system.  The companion will lose mass as material flows from its outer atmosphere onto the WD.  Using Equation 9, $\dot{M}=10^{-8}$ M$_{\odot}$ yr$^{-1}$ from 2004, and the assumption that the current decline rate continues, we estimate $M_{comp}$ will decrease by $6\times 10^{-6}$ M$_{\odot}$.  This number has considerable uncertainty; for example, the number would be a factor of six times larger had we adopted the accretion rate measured from the observed period change presumably caused by conservative mass transfer (Patterson et al. 1998).  The change in mass for the WD from just before the `1866' event until accretion stops is dominated by how much mass was ejected in the `1866' explosion.  The mass ratio before the `1866' nova is approximately $q=M_{comp}/M_{wd}\approx0.12/1.3=0.092$, and this ratio will get smaller by -0.0000025 over the duration of the RN state.
	
	Six mechanisms are operating to change the orbital period of the system during the RN phase:  (1) The mass lost during eruptions will cause the orbital period to increase.  Schaefer \& Patterson (1983) derive that the period change is
\begin{equation}
\Delta P = A P_{orb} M_{shell} M_{comp}^{-1}, 
\end{equation}
where for T Pyx the orbital period is $P_{orb}=1.83$ hours, the ejected shell mass is $M_{shell}$, and the mass of the companion star is $M_{comp}=0.12$ M$_{\odot}$ (Selvelli et al. 2008; Knigge et al. 2000).  The parameter $A$ depends on the specific angular momentum of the ejecta, which we assume to equal the specific angular momentum of the white dwarf, and the fraction of ejecta that is captured by the companion star, which we assume to equal the fractional coverage of the sky subtended by the companion star (1.0\%), for $A=2.13$.  For the big `1866' event, $M_{shell}\sim 10^{-4.5}$ M$_{\odot}$.  With this, the orbital period will increase by +0.0010 hours, with a substantial uncertainty.  (2) Each RN event will also cause mass loss to the system, which will increase the orbital period.  For one eruption, with $M_{shell}\sim 2 \times 10^{-6}$ M$_{\odot}$, we get $\Delta P = 0.000064$ hours, with substantial uncertainty.  For the five known RN events, the total period increase from this mechanism is roughly +0.00005 hours.  (3) During the nova eruptions, as the companion star passes through the expanding nova shell, frictional angular momentum losses will occur during this common envelope phase, and the orbital period will decrease.  Kato \& Hachisu (1991) and Livio (1991) estimated the size of this effect for T Pyx in particular, but this should be re-evaluated with our better input values.  With equations from Kato \& Hachisu (1991, Equations 1-4), Livio (1991, Equation 5), and Frank et al. (2002, Equation 4.18), we find
\begin{equation}
\Delta P = -0.75 P_{orb} (M_{shell}/M_{comp}) (V_{orb}/V_{exp}) (R_{comp}/a)^2,
\end{equation}
where the orbital velocity of the companion star is $V_{orb}$, the semi-major axis is $a$, and the radius of the companion star is $R_{comp}$.  For the big `1866' nova eruption with $M_{shell}\sim 10^{-4.5}$ M$_{\odot}$ and $V_{exp}=600$ km s$^{-1}$, we find that the orbital period will decrease by -0.000013 hours.  (4) Similarly, the RN events will also see a period decrease due to the frictional angular momentum loses.  For $M_{shell}\sim 2 \times 10^{-6}$ M$_{\odot}$ and $V_{exp}=2000$ km s$^{-1}$, we find that the period decreases by 250 billionths of an hour.  For five such RN eruptions, the total period change is -0.0000012 hours.
(5) Throughout the entire duration of the recurrent nova state, $\Delta T \sim 200$ years, ordinary gravitational radiation will carry away angular momentum and drive the orbital period down.  From Rappaport et al. (1982, Equation 19) and Frank et al. (2002, Equation 4.18), we can calculate the period change as
\begin{equation}
\Delta P = -860 G^{5/3} c^{-5} \Delta T (4/3 - q)^{-1} (M_{comp}+M_{wd})^{-1/3} M_{comp} M_{wd} P_{orb}^{-5/3},
\end{equation}
where the mass ratio is $q=M_{comp}/M_{wd}\approx0.092$.  For T Pyx, the change in period is -0.00000011 hours.  (6) Ordinary conservative mass transfer in the quiescent system during the recurrent nova state will cause the orbital period to increase.  From Frank et al. (2002, Equations 4.14 and 4.18), we get 
\begin{equation}
\Delta P = -3  P_{orb} (1 - q) \Delta T \dot{M} M_{comp}^{-1}.
\end{equation}
For T Pyx in its recurrent nova state, we estimate that the integral over time of $\Delta T \dot{M}$ is $6 \times 10^{-6}$ M$_{\odot}$.  This leads to a period increase of roughly +0.00025 hours.

	We can estimate the total change in the orbital period during the RN state by adding up the six contributions from the preceding paragraph.  We find that the orbital period {\it increases} by +0.0013 hours.  This is necessarily an approximate answer, because the various masses and the average accretion rates are not accurately known.  All but two of the mechanisms are negligible in size, so that only the effects of the mass loss during the `1866' outburst and the conservative mass transfer throughout the entire RN state are significant contributors.  Both of these dominant mechanisms act to {\it increase} the period.  Thus, T Pyx will have its orbital period increase by about a part-per-thousand from 1866 into the near future.
	
	The semimajor axis, $a$, must change from its current value of roughly 0.86 R$_{\odot}$.  With the changes described above and Equation 4.2 of Frank et al. (2002), we find that $a$ will increase by 0.00039 R$_{\odot}$ during the RN state, a fractional change of $0.046\%$.  The change in orbital period and separation will lead to a change in the size of the Roche lobe as well.  With the changes described above and Equation 4.6 of Frank et al. (2002), we find that the change in $R_{Roche}$ will be an increase of 0.000077 R$_{\odot}$.
	
	\subsection{The Hibernation State}
	
	The accretion in T Pyx has never been sufficient to fully sustain the supersoft source, so the accretion rate has steadily declined.  We know of no reason why this situation should stop now that the accretion rate is far below the maximum.  $\dot{M}$ will keep falling, the heating of the companion star will turn off, and the companion star will return to its previous non-irradiated state.  This normal size of the companion star will be close to the size just before the `1866' event, but at the end of the RN state, the size of the Roche lobe will have significantly expanded.  While the relaxed companion star was in contact with its Roche lobe in 1865, it will be out of contact with its Roche lobe in 2065.  That is, when the RN state is over, there will be effectively zero accretion.  T Pyx will go into a state of hibernation, where we only see the faint light from the companion and the WD, with no accretion disk.
	
	Unsurprisingly, the actual situation is slightly more complex.  We must account for the slight shrinkage of the companion star due to its mass loss during the RN phase, while realizing that some accretion will occur even when the companion star is out of contact with its Roche lobe due to the exponential scale height of its atmosphere.  We calculate the amount the companion shrinks by using the mass-radius relation for low mass stars: $R_{comp} \propto M_{comp}^{0.8}$ (S\'egransan et al. 2003; Ribas 2006).  (Recall that we are comparing the star at two times when it is not puffed up due to irradiation.)  The companion star will shrink in radius by 5 km due to the loss of $\sim 6 \times 10^{-6}$ M$_{\odot}$ during the RN phase.  When combined with the expansion of the Roche lobe, by 54 km, we see that the surface of the companion star has moved to 59 km inside its Roche lobe.
	
	The scale height of the atmosphere of the companion star is 
\begin{equation}
H_*=kT_*/g\mu m_H, 
\end{equation}
where $k$ is the Boltzmann constant, $T_*$ is the surface temperature (roughly 3000 K), $g$ is the surface gravity of $1.14\times 10^5$ cm s$^{-2}$, $\mu$ is the mean atomic weight for the composition of the secondary star (roughly 1.4 for solar abundances), and $m_H$ is the mass of hydrogen.  For T Pyx, we calculate the scale height is 12 km.

	The accretion rate will be proportional to $e^{-\Delta R/H_*}$, where $\Delta R$ is the distance from the surface of the star to the Roche lobe (Frank et al. 2002, Equation 4.19).  Just before the `1866' eruption, when the companion star was in its normal state with no irradiation, the accretion rate was that associated with the loss of gravitational radiation, $\dot{M}=4 \times 10^{-11}$ M$_{\odot}$ yr$^{-1}$.  We take the star surface at this time to be at the Roche lobe, so $\Delta R=0$.  At the start of the hibernation state, the star has again relaxed to its normal state with no irradiation, except now the expansion of the Roche lobe and the contraction of the star make $\Delta R=59$ km.  With the scale height of 12 km, the accretion will be lower by a factor of 140.  Thus, during hibernation, the expected accretion rate has not dropped completely to zero, but is instead around $\dot{M}=3 \times 10^{-13}$ M$_{\odot}$ yr$^{-1}$.
	
	During hibernation, the system will be faint.  The 0.12 M$_{\odot}$ secondary star should have an absolute magnitude of around 15 mag, which at a distance of 3,500 pc corresponds to an apparent magnitude of around 28 mag.  The magnitude of the WD is highly uncertain because we do not know its temperature, but given that its surface area is small (because its mass is near the Chandrasekhar mass) we expect that the WD will not dominate the brightness.  From Equation 9, and scaling from the approximate accretion rate around the year 2000 ($\dot{M}=10^{-8}$ M$_{\odot}$ yr$^{-1}$), we can estimate the B-band magnitude as
\begin{equation}
B=15.5 - 1.25 \times \log (\dot{M}/10^{-8}).
\end{equation}
The residual disk is thus around 21 mag during hibernation.  This feeble disk will dominate over both the red and the white dwarf.
	
	In this hibernation state, only two mechanisms will operate to change the orbital period of T Pyx.  The gravitational radiation will still relentlessly grind down the period with $\dot{P}=-5.3 \times 10^{-10}$ hours per year.  The conservative mass transfer will increase the orbital period by $\dot{P}=+1.2 \times 10^{-11}$ hours per year.  Thus, the gravitational radiation dominates by a factor of 45.  
	
	We must ascertain the length of the hibernation phase, $\Delta T_{hib}$.  The net period derivative of T Pyx in hibernation will be $\dot{P}=-5.2 \times 10^{-10}$ hours per year.  From Equations 4.6 and 4.18 from Frank et al. (2002), we have that $\Delta P_{hib}/P = 1.5 \Delta R_{Roche}/R_{Roche}$, where the differences are across the entire time interval of the hibernation state (with $\Delta P_{hib} = \dot{P} \Delta T_{hib}$).  With this, 
\begin{equation}
\Delta T_{hib} = 1.5 (\Delta R_{Roche}/R_{Roche}) (P/\dot{P}). 
\end{equation}	
We have already estimated that the Roche lobe must contract by $\Delta R_{Roche}=59$ km to bring the system back into the contact as in its ordinary CV state.  We thus calculate that the hibernation state will last for 2.6 million years.

	\subsection{The Ordinary CV State}
	
	Eventually, gravitational radiation will drive the angular momentum loss such that the hibernating T Pyx system will  return to contact.  The secondary will make contact with its Roche lobe over an extended time due to the exponential scale height of its atmosphere.  We take the beginning of the ordinary CV state to be when the accretion rate is back to the steady state value associated with gravitational radiation losses.  Once in this state, the binary will look like all other CVs with an orbital period below the period gap.  Its light will be dominated by the accretion disk, with a mild accretion rate of $\dot{M}=4 \times 10^{-11}$ M$_{\odot}$ yr$^{-1}$, which would make the system around B=18.5 mag.  Such systems would likely never be discovered with current observing programs.
	
	T Pyx will remain a faint contact binary with a low accretion rate for a long time.  The accretion disk will be the dominant light source.  At its large distance, no one would recognize T Pyx as the unique system that it now appears to be.  Given their low luminosity, our galaxy might well have large numbers of such systems hidden away; only the closest examples have any chance of being recognized as CVs.  
	
	We can calculate the length of this ordinary CV state.  We know from nova trigger theory (e.g., Yaron et al. 2005) that a high-mass WD accreting at the rate of $4 \times 10^{-11}$ M$_{\odot}$ yr$^{-1}$ will require a trigger mass of $3 \times 10^{-5}$ M$_{\odot}$.  As the accretion rate remains constant, we derive a time interval of 750,000 years required to accumulate the trigger mass on the white dwarf.  This does not account for the unburned material accreted during the recurrent nova state and the hibernation state after 1967, however this amount of material is small compared to the trigger mass, so it is negligible.  As such, with reasonable confidence and accuracy, we know that the ordinary CV state of T Pyx will last about three-quarters of a million years.  When the trigger mass has accumulated, T Pyx will undergo another ordinary nova eruption (just like the `1866' eruption), beginning the cycle again.  The full cycle, which lasts approximately 3,350,200 years, can be seen in graphical form in Figures 10 and 11, in terms of magnitude and accretion rate, respectively.

\section{Stars Like T Pyx}

	We have made the point that T Pyx represents a rare case (a very high-mass WD and a very close companion star) currently in a rare state (lasting only a century or so out of every cycle lasting $\sim3,300,000$ years).  As such, it is fair to expect that T Pyx should be unique amongst known stars.  Nevertheless, it is useful to seek analogous systems for the various stages in T Pyx's cycle.
	
	The ten known galactic RNe are the only novae systems which can be studied in detail both before and after eruption.  Of these ten, eight have evolved companions; the expansion of the companion star drives the accretion and evolution in these systems.  T Pyx and IM Nor are the other two known galactic RNe.  As we have shown, the high accretion rate in T Pyx is apparently driven by a continuing - but not self-sustaining - supersoft source.  Little is known about IM Nor; it has an orbital period of 2.45 hours, right in the middle of the period gap, but we do not know whether it has a secular decline, a nova shell, or a supersoft source.  Although there is no published estimate of IM Nor's accretion rate, we know that it must be high to support its short recurrence time scale, certainly higher than possible in ordinary CVs (Patterson 1984).  It is possible that the accretion in IM Nor is driven by a mechanism similar to that in T Pyx.

	For T Pyx, we have learned that the RN state started with an ordinary nova outburst that was preceded by a low accretion rate and followed by a high accretion rate.  That is, T Pyx has an estimated change in quiescent brightness of around 5.0 mag (i.e., a factor of 100 in brightness) from before the `1866' eruption to after.  Perhaps we can recognize systems like T Pyx by this large change in magnitude.  The obvious candidate is V1500 Cyg, a very fast naked eye nova from 1975.  This system was positively identified on two very old plates, so $B=21.5$ before eruption (Duerbeck 1987; Wade 1987).  After eruption, the light curve has gone to quiescence at around seventeenth mag (and declining to around nineteenth mag in recent years).  This is a difference of $\gtrsim2.5$ mag (i.e., by over a factor of 10 in brightness) from pre-eruption to post-eruption.  Clearly, something substantial had to change in the system, likely due to the 1975 eruption.  Significantly, V1500 Cyg is well known to be a strong supersoft source, with the best evidence being that the hemisphere facing towards the WD has a temperature of 8000K, while the backside hemisphere has a temperature of 3000K as appropriate for a 0.29 M$_{\odot}$ star (Somers \& Naylor 1999).  Just as with T Pyx, the supersoft source in V1500 Cyg is declining with time; over twenty years the supersoft flux has been falling roughly as the time since the eruption to the -1 power, with a prediction that the source will turn off in roughly two centuries.  So V1500 Cyg appears to be similar to T Pyx in having an ordinary nova eruption in a short-period system inducing a bright-but-fading supersoft source that powers a sudden jump in the accretion rate.  Nevertheless, V1500 Cyg has substantially different properties from T Pyx, with it being a polar and having $M_{WD}\sim1.15$ M$_{\odot}$ (Hachisu \& Kato 2006).
	
	Our group in Louisiana has been systematically measuring pre-eruption magnitudes of old novae by direct examination of archival photographic plates at both Harvard College Observatory and Sonneberg Observatory.  Almost all novae have the pre-eruption magnitude equal to the post-eruption magnitude to within measurement uncertainties and the usual flickering and variability of quiescent novae.  However, we find six other novae with a large pre-to-post magnitude difference (Collazzi et al. 2009; Schaefer \& Collazzi 2009).  For example, V723 Cas stands out as having a large brightening across the eruption.  The Palomar Sky Survey shows the magnitude to be $B=18.76\pm0.3$, while Goranskij et al. (2007) and the {\it AAVSO} light curves show that the nova has stayed constant at $B=15.75\pm0.5$ for the last decade.  With a difference of $3.0\pm0.6$ mag (i.e., by a factor of $>10$ in brightness), we have a clear case of a nova event changing the accretion level.  V723 Cas is one of the few CNe that has turned on a long-lasting strong supersoft source at the time of its eruption, with this source lasting for at least twelve years with no signs of turning off (Ness et al. 2008).  But again, other aspects of V723 Cas are different from T Pyx, e.g., the orbital period is 16.6 hours (Goranskij et al. 2007) and the ejected material is neon-rich (Iijima 2006).
	
	RW UMi erupted in 1956, was at $B>21$ mag on the first epoch Palomar Sky Survey plate from 1952, and yet was at B=18.33 in 1989 on the second epoch Palomar Sky Survey.  RW UMi has changed its quiescent brightness by $>2.67$ mag (i.e., over a factor of 10 in brightness) from pre- to post-eruption.  RW UMi is like T Pyx in a number of other ways: it has a very short orbital period of 1.42 hours (Retter \& Lipkin 2001), it is slowly declining in optical brightness at the rate of 0.02 mag yr$^{-1}$ (Bianchini et al. 2003), and it has a nova shell (Cohen 1985; Esenoglu et al. 2000).  This makes RW UMi the fourth example of a significant and sustained increase in accretion across a classical nova eruption.  But RW UMi is different from T Pyx in that it is likely an intermediate polar (Bianchini et al. 2003).
	
	We now have seven examples of novae that behaved like T Pyx in that the quiescent brightness (and thus the accretion rate) increased by a large factor across a classical nova event.  The brightness increase of T Pyx has solid precedent, and the cause is apparently due to the initiation of a long-lasting supersoft source (Schaefer \& Collazzi 2009).  Apparently, the conditions required to create the sustained supersoft source vary considerably, such that moderate mass WDs, highly magnetized WDs, and systems with moderately-long orbital periods can also initiate long intervals of nuclear burning on their surfaces.

\section{T Pyx as a Supernova}
	
	Many papers have pondered whether T Pyx will ultimately collapse as a Type Ia supernova.  Anupama (2002), Hachisu et al. (2008), and Livio \& Truran (1992) all conclude that T Pyx will become a Type Ia supernova, always based on the general consideration that the WD is near the Chandrasekhar mass and accreting material at a high rate.  But these papers are either not particular to T Pyx or are without detailed calculations.  Knigge et al. (2000) considered that T Pyx {\it might} become a supernova, provided that the combined mass of the two stars is above the Chandrasekhar limit.  But they did not know that T Pyx has undergone a large secular decline, that the supersoft source is not self-sustaining (Schaefer 2005), and that the system is not now a supersoft source (Selvelli et al. 2008), so their conclusion on the fate of T Pyx does not still hold.  Selvelli et al. (2008) concluded that T Pyx would not become a supernova, with the primary reason being their estimates of a high ejected mass and an accreted mass that could not be anywhere near as large, so the white dwarf could not be gaining mass.  But as we have seen (Section 7.2), their high $M_{ejecta}$ cannot be right, as they were aware, so their conclusion on the fate of T Pyx does not follow.  All previous speculation on the fate of T Pyx is now irrelevant because the dominant event (the high-mass `1866' eruption) was not known or considered.
	
	A good indication that T Pyx cannot become a supernova is simply that it does not have the required total mass.  For our preferred WD mass of 1.25 M$_{\odot}$ and the fairly well-determined companion star mass of 0.12 M$_{\odot}$, the system does not exceed the Chandrasekhar mass and cannot collapse.  This argument is easy to overcome as it is possible that the WD is at 1.30 M$_{\odot}$ or more, allowing for a collapse.  But this does not take into account the fact that most of the material accreting onto the WD gets blown off in classical nova events.  Yaron et al. (2005) calculate that the nova eruptions will eject more mass than is being accreted, in which case the system can never exceed the Chandrasekhar mass.  Even if theory is wrong, and, say, half the accreted material is retained by the WD, then T Pyx can become a supernova only if the WD is within 0.06 M$_{\odot}$ of the limit, and this does not seem likely given the results from Section 7.3.  With this, we already know that there is little real chance of T Pyx ever collapsing as a supernova.
	
	{\it If} the system has enough mass and if the WD is gaining mass, then it will almost certainly reach the Chandrasekhar mass and will produce a Type Ia supernova.  The key question is whether the WD is accreting more mass than it is ejecting.  The RN events themselves involve so little mass (and occur only 6 or 7 times every 3.3 million years) that they are negligible in comparison to the `1866' eruption.  We do not have any real measure of the accreted mass (say, from the $\dot{M}$ and the time interval between eruptions) for the `1866' event; the best we can do is equate this to the trigger mass within some physical model of nova eruptions.  We must therefore answer the question of whether the mass ejected in `1866' is more or less than the trigger mass for that eruption.
	
	We have three arguments pointing to different conclusions.  First, nova models indicate that events like the `1866' nova should eject more mass than is accreted between eruptions.  For $M_{WD}$ equal to 1.25 or 1.4 M$_{\odot}$ and $\dot{M}\sim4\times10^{-11}$ M$_{\odot}$ yr$^{-1}$, the ejected mass is larger than the accreted mass by 15-35\% (Yaron et al. 2005).  The long inter-eruption interval allows for the hydrogen deposited on the WD surface to diffuse into the older material so that this older material will be lifted off when the thermonuclear runaway eventually happens.  The implication is that the WD is losing mass, and so no collapse will occur.  Second, the `1866' ejecta (i.e., the now visible knots in the shell) have metal abundances that are consistent with solar abundance (Contini \& Prialnik 1997; Williams 1982), which implies that there can be no significant amount of dredged-up material.  To be quantitative, Yaron et al. (2005) give the metalicity ($Z$) of the ejecta from theory as being from 0.18-0.33, whereas the observations imply $Z\approx0.02$.  We can offer no explanation for how this high-metalicity material can become diluted or otherwise appear as solar-metalicity material.  For example, to dilute the ejected material down to the observed metallicity, say, from the interstellar medium, it would take roughly one order of magnitude more mass swept up than was originally in the knot, and this has not occurred because the knots would have decelerated at greatly different rates so as to cause a wide dispersion in Figure 6 (see Section 4.4).  With no material from below the layer of accreted material, the ejected material can only be less than the accreted mass, and the WD must be gaining mass over each cycle.  Third, if the WD is losing mass and it must now be near the Chandrasekhar mass, then with a simple extrapolation of this situation backwards in time, the WD has recently been more massive than the Chandrasekhar limit, and this is not possible.  An answer is that the WD in T Pyx {\it formed} with a near maximal mass and has been slowly whittled away ever since.  In all, we have a conflict between the expectations of theoretical models versus the observations.  We do not know how to resolve this conflict with any useable confidence.
	
	We have already made estimates of the mass ejected and the trigger mass.  From Section 7.1, we found $M_{ejecta}=3\times10^{-5} ~d_{3500}^2$ M$_{\odot}$, with an uncertainty of perhaps a factor of two or three.  For WD masses from 1.2-1.4 M$_{\odot}$ and for accretion rates from $1-10 \times 10^{-11}$ M$_{\odot}$ yr$^{-1}$, the trigger mass varies from $1-30 \times 10^{-5}$ M$_{\odot}$ with a best estimate of $3\times10^{-5}$ M$_{\odot}$.  So for our best values, the trigger mass equals the ejecta mass, and the WD is neither gaining nor losing mass.  But the possible range of mass gained by the WD goes from -4 to +28 times $10^{-5}$ M$_{\odot}$.  Undoubtedly, arguments will be made to narrow the range of this estimate, perhaps substantially, but we do not see that any confident case can be made that will reduce the error bars to the point where a decision can be presented as to whether the WD is increasing its mass.
			
	If T Pyx does in fact have a WD that increases in mass, the time scale for it to become a Type Ia supernova is highly uncertain because of our previously-mentioned lack of knowledge about the mass of the WD and how it changes.  Nevertheless, we can make plausible estimates and see how things change over the range of inputs.  We adopt the ranges of $1-10 \times 10^{-11}$ M$_{\odot}$ yr$^{-1}$ for $\dot{M}$ and 1.2-1.4 M$_{\odot}$ for $M_{WD}$, with our best estimates of $4 \times 10^{-11}$ M$_{\odot}$ yr$^{-1}$ and 1.25-1.30 M$_{\odot}$ respectively.  For any particular choices in this range, we can use theoretical models to get the trigger mass $M_{trig}$.  The duration of the ordinary CV state will be $M_{trig}/\dot{M}$, and this can be added to 2,600,200 years to get the total cycle duration.  We assume that the total mass accreted will be equal $M_{trig}$ and that the ejected mass will equal $\epsilon M_{trig}$, where $\epsilon$ is an adjustable parameter that must be positive (to have some ejecta) and not greater than unity (to avoid dredge-up since it is not seen in the abundances).  The mass gain during each cycle is then $(1-\epsilon)M_{trig}$.  With this, we can track the mass of the WD and the ejected mass through time.

	With our best-estimate values for $M_{WD}$ and $\dot{M}$, the time to collapse will only depend on $\epsilon$.  As $\epsilon$ rises from 0.0 to 0.9, the time until T Pyx goes supernova ($T_{SN}$) rises from 6.3 to 63 billion years.  The time is longer than the age of the Universe for $\epsilon$ larger than 0.5.  But any low value of $\epsilon$ is already refuted by our knowledge that $M_{ejecta}\sim 10^{-4.5} d_{3500}^2$ M$_{\odot}$.  We cannot push $M_{ejecta}$ below something like $10^{-5}$ M$_{\odot}$.  This constraint forces some combination of low $M_{WD}$ and high $\epsilon$, both of which will force long $T_{SN}$.  For $M_{WD}=1.35$ M$_{\odot}$ this forces $\epsilon>0.35$ and $T_{SN}=9.7$ billion years, while for $M_{WD}=1.25$ M$_{\odot}$ this gives $\epsilon>0.40$ and $T_{SN}=22.5$ billion years.  Given the normal solar metalicity of the ejecta, a system age of 9.7 billion years is not plausible.  T Pyx is unlikely to become a supernova anytime in the next many billion years.  
	
	A more important point is that the lifetime of the state leading up to any supernova collapse is comparable to the age of the Universe (and then only for optimal parameters), so the death rate of such systems must be incredibly low.  The death rate by collapse (i.e., the Type Ia supernova rate from this class of systems) will be the number of systems in our galaxy divided by the lifetime of the systems.  This rate is to be compared to $0.3\pm0.2$ Type Ia supernovae per century in our Milky Way (Wheeler \& Benetti 2000).  Given that the space density of {\it all} CVs is $\sim 1 \times10^{-5}$ pc$^{-3}$ for the disk population (Patterson 1984; Pretorius et al. 2007; Rogel et al. 2008), the effective volume for the disk with a scale height of 150 pc is less than $10^{11}$ pc$^{3}$, and the rough equality in numbers of the bulge and disk populations of CVs (Shafter 2002), then the total number of CVs in our Milky Way must be less than two million or so.  Not all CVs are systems like T Pyx, as it is rare to have a near-Chandrasekhar mass WD, so we can set an upper limit that the number of such systems must be less than 100,000 or so.  With this, the average lifetime of the system must be roughly 30 million years for the T Pyx systems to contribute all of the observed supernova rate for our Milky Way.  The only way to achieve this sort of a lifetime is if the T Pyx-like system was formed with a WD mass much closer than 0.001 M$_{\odot}$ to the Chandrasekhar mass.  This strong result forces us to conclude that systems like T Pyx cannot provide any significant contribution to the observed Type Ia supernova rate.
	
	A question then arises as to the progenitor of T Pyx.  That is, given that the T Pyx WD now has a mass near the Chandrasekhar mass, then how did the WD get this massive given that the time required to raise its mass through the RN range ($\sim 1.2-1.4$ M$_{\odot}$) is greater than the age of the Universe.  For example, for  $\dot{M}=4 \times 10^{-11}$ M$_{\odot}$ yr$^{-1}$ and $\epsilon=0.35$, the time for when the WD mass rises from 1.2 M$_{\odot}$ until it gets to 1.35 M$_{\odot}$ is 16.4 billion years.  We cannot produce T Pyx by simply adding mass to a low-mass WD through our three part cycle.  The only alternative is that the WD formed with high mass, either as the core of a near-maximal main sequence star or as a more usual WD with matter piled on during some common envelope phase.
	
	In all, we strongly conclude that T Pyx will not become a Type Ia supernova within many Hubble times if ever, and that systems like T Pyx cannot provide even a part per thousand of the observed supernova rate.
	
\section{Conclusions}

	We have used our new {\it HST} images, with a long time baseline of $2007.5-1994.1=13.4$ years, to measure the changes in the T Pyx shell.  From this, we can derive the surprising properties of the knots and deduce the circumstances of their origin.  When combined with the secular decline in brightness over the last 120 years, we get a clear picture of the evolutionary cycle of T Pyx.  Here are our conclusions:
	
	(1) We find that the knots are expanding radially with a typical velocity of 600 km s$^{-1}$ and a range of velocities of 500-715 km s$^{-1}$, for an assumed distance of 3,500 pc.  This expansion is proportional to the distance of the knot from T Pyx (i.e., it is homologous), with all of the knot radial distances in 1995.8 being a factor of $0.917\pm0.003$ times smaller than in 2007.5, and the knot radial distances in 1994.1 being a factor of $0.907\pm0.005$ times smaller than in 2007.5.  All models for which the knots undergo some significant deceleration must have a large scatter in the fractional expansion of the knots, so we conclude that no significant deceleration has occurred.  With this, a simple extrapolation back in time is correct, giving an ejection date for the knots of $1866\pm5$.
	
	(2) We find that some knots have turned on in our 2007.5 image, while others have faded by typically 1.5 mag since 1995.8.  The late turn-on of many knots demonstrates that the knots must be powered by collisional shocks.  The shell in the 1970s and 1980s was powered by collisions of  the ejecta from the 1944 eruption with the `1866' knots; since then, the knots are being powered by collisions with the ejecta from the 1967 eruption.  The fading rate of the knots gives a rough estimate that the electron density is $\sim$6000 cm$^{-3}$.  With corrections for the turn-on and fading of the knots, we estimate the mass in the knots to be $\sim$10$^{-4.5}$ M$_{\odot}$.
	
	(3) The lighter and faster ejecta from later RN outbursts create a wind pushing into the `1866' shell.  In this case, we know that Rayleigh-Taylor instabilities will fragment the `1866' shell into knots, just as we see with {\it HST}.  The fast RN ejecta will push through and form the smooth and faint outer halo.
	
	(4) The ejection velocity and mass of the `1866' event are comparable to those of ordinary nova eruptions, whereas they are completely different from the later RN events.  Indeed, the only way to get such a high mass ejected from a high-mass WD is to have the material added at a very low accretion rate.  Such a low accretion rate is exactly what is expected for T Pyx in the ordinary CV case of a system below the period gap being driven only by gravitational radiation.  Before it was in its RN state, T Pyx must have been an ordinary CV with $\dot{M}\sim4\times10^{-11}$ M$_{\odot}$ yr$^{-1}$ accumulating matter for around 750,000 years so as to produce an ordinary nova event around 1866.
	
	(5) Some singular event put T Pyx into a state of high accretion around the year 1866, and the only good idea is that the white dwarf became a supersoft source which drove the high $\dot{M}$.  A reasonable scenario is that the normal supersoft source present towards the end of ordinary nova eruptions was extended indefinitely due to the near-Chandrasekhar $M_{WD}$ and the proximity of the companion star, driving enough wind to sustain the nuclear burning on the surface of the WD.  The century-long secular decline in brightness demonstrates that the accretion is turning off (with $\dot{M}$ in 2009 at 3\% of its value in 1890), so the supersoft source is not perfectly self-sustaining and indeed is not now visible even with {\it XMM-Newton}.  Calculations of the size of the companion star with respect to its Roche lobe size demonstrate that the accretion will soon fall to $\dot{M}\sim3\times10^{-13}$ M$_{\odot}$ yr$^{-1}$.  With this, T Pyx will fade to something like 21 mag, which puts it in a state of hibernation.
	
	(6) T Pyx is undergoing a multi-part evolutionary cycle (see Figures 10 and 11).  This cycle starts out with T Pyx looking like an ordinary CV below the period gap, in a state which lasts roughly 750,000 years.  When enough matter has accumulated on the WD, T Pyx undergoes an ordinary nova outburst, which then starts a supersoft source, a high accretion rate, and RN eruptions.  This RN state slowly winds down as the accretion turns off nearly completely, with no further eruptions expected in the foreseeable future.  With the accretion turned off, T Pyx enters its hibernation state.  Inevitably, gravitational radiation will drive the system back into contact and T Pyx will re-enter its ordinary CV state, at which point the cycle starts over again.
	
	(7) Whether T Pyx will become a Type Ia supernova depends on whether the WD is gaining or losing mass throughout this evolutionary cycle, and that depends only on whether the ordinary nova event has $M_{ejecta}$ greater than its trigger mass.  The best estimates place the ejection mass equal to the trigger mass (so that the WD neither gains nor loses mass over each eruption cycle), but the moderately large uncertainties mean that the change in the WD mass could be either positive or negative.  Indeed, given the best estimate of the mass of the WD (1.25 M$_{\odot}$), the good estimate for the mass of the companion star (0.12 M$_{\odot}$), and the large fraction of the trigger mass ejected every eruption, there is not enough mass in the system for T Pyx to {\it ever} collapse as a supernova.  For systems like T Pyx, even if the WD is gaining mass, the long cycle time suggests that it will take longer than the age of the Universe to ultimately collapse.  Thus, T Pyx and stars like it cannot provide any significant rate of Type Ia supernovae.
	
	~
	
	We thank {\it NASA} and {\it HST} for a grant to help support this work.

\clearpage

\begin{deluxetable}{llllllllllllllll}
\tabletypesize{\scriptsize}
\rotate
\tablecaption{Knot Summary
\label{tbl1}}
\tablewidth{0pt}
\tablehead{
\colhead{Knot}   &
\colhead{$X-X_T$}   &
\colhead{$Y-Y_T$}   &
\colhead{$\theta_{2007}$}  &
\colhead{$R_{2007}$}   &
\colhead{$\Delta m_{07}$}   &
\colhead{$\theta_{1995}$}  &
\colhead{$R_{1995}$}   &
\colhead{$\Delta m_{95}$}   &
\colhead{$\theta_{1994}$}  &
\colhead{$R_{1994}$}   &
\colhead{$\Delta m_{94}$}   &
\colhead{$R_{95}/R_{07}$}   &
\colhead{$R_{94}/R_{07}$}   &
\colhead{$\Delta m_{07-95}$}   &
\colhead{$\Delta m_{95-94}$}
}
\startdata

K1	&	-13.3	&	14.0	&	-44	&	19.3	&	5.61	&	-41	&	18.4	&	6.41	&	-30	&	15.3	&	5.97	&	0.953	&	0.793	&	-0.80	&	0.44	\\
K2	&	-17.0	&	24.9	&	-34	&	30.1	&	6.03	&	-34	&	28.0	&	5.64	&	-34	&	27.4	&	5.77	&	0.930	&	0.910	&	0.39	&	-0.13	\\
K3	&	24.1	&	-21.0	&	131	&	32.0	&	5.76	&	131	&	28.4	&	4.94	&	131	&	28.0	&	4.97	&	0.888	&	0.875	&	0.82	&	-0.03	\\
K4	&	-23.5	&	23.4	&	-45	&	33.1	&	6.03	&	-45	&	30.6	&	5.15	&	-45	&	30.6	&	5.31	&	0.924	&	0.924	&	0.88	&	-0.16	\\
K5	&	-57.0	&	0.3	&	-90	&	57.0	&	5.56	&	-89	&	51.7	&	6.64	&	-88	&	51.0	&	6.78	&	0.907	&	0.895	&	-1.08	&	-0.14	\\
K6	&	12.1	&	-58.1	&	168	&	59.3	&	5.73	&	168	&	53.5	&	4.64	&	168	&	52.9	&	5.41	&	0.902	&	0.892	&	1.09	&	-0.77	\\
K7	&	-25.4	&	-58.2	&	-156	&	63.5	&	4.84	&	\ldots	&	\ldots	&	\ldots	&	\ldots	&	\ldots	&	\ldots	&	\ldots	&	\ldots	&	\ldots	&	\ldots	\\
K8	&	71.8	&	-2.4	&	92	&	71.8	&	4.77	&	93	&	66.4	&	4.44	&	93	&	65.8	&	4.39	&	0.925	&	0.916	&	0.33	&	0.05	\\
K9	&	72.5	&	-8.1	&	96	&	73.0	&	4.77	&	\ldots	&	\ldots	&	\ldots	&	\ldots	&	\ldots	&	\ldots	&	\ldots	&	\ldots	&	\ldots	&	\ldots	\\
K10	&	71.7	&	-23.4	&	108	&	75.5	&	5.06	&	107	&	69.9	&	4.67	&	107	&	69.5	&	4.67	&	0.926	&	0.921	&	0.39	&	0.00	\\
K11	&	-39.2	&	69.7	&	-29	&	80.0	&	5.72	&	\ldots	&	\ldots	&	\ldots	&	\ldots	&	\ldots	&	\ldots	&	\ldots	&	\ldots	&	\ldots	&	\ldots	\\
K12	&	85.5	&	8.2	&	85	&	85.9	&	4.98	&	84	&	78.4	&	4.61	&	84	&	77.5	&	4.64	&	0.913	&	0.902	&	0.37	&	-0.03	\\
K13	&	82.6	&	-24.4	&	106	&	86.1	&	6.41	&	106	&	80.3	&	6.52	&	106	&	79.6	&	6.26	&	0.933	&	0.925	&	-0.11	&	0.26	\\
K14	&	85.9	&	-8.1	&	95	&	86.3	&	5.07	&	95	&	79.3	&	5.34	&	95	&	78.6	&	5.28	&	0.919	&	0.911	&	-0.27	&	0.06	\\
K15	&	72.9	&	-53.0	&	126	&	90.2	&	5.52	&	126	&	81.8	&	4.31	&	125	&	81.7	&	7.32	&	0.907	&	0.906	&	1.21	&	-3.01	\\
K16	&	66.3	&	-67.8	&	136	&	94.8	&	5.95	&	135	&	86.6	&	6.90	&	135	&	85.9	&	6.70	&	0.914	&	0.906	&	-0.95	&	0.20	\\
K17	&	53.6	&	80.8	&	34	&	96.9	&	5.49	&	33	&	88.0	&	6.90	&	34	&	87.4	&	6.51	&	0.908	&	0.902	&	-1.41	&	0.39	\\
K18	&	64.5	&	81.7	&	38	&	104.1	&	5.00	&	39	&	98.4	&	6.18	&	39	&	97.7	&	6.00	&	0.945	&	0.939	&	-1.19	&	0.18	\\
K19	&	-10.8	&	-103.9	&	-174	&	104.4	&	5.44	&	-174	&	96.4	&	6.47	&	-174	&	95.4	&	6.18	&	0.923	&	0.914	&	-1.03	&	0.29	\\
K20	&	-82.9	&	70.8	&	-50	&	109.0	&	5.53	&	-50	&	101.5	&	5.28	&	-50	&	100.5	&	4.94	&	0.931	&	0.922	&	0.25	&	0.34	\\
K21	&	-107.6	&	-18.6	&	-100	&	109.2	&	5.28	&	-100	&	99.1	&	3.69	&	\ldots	&	\ldots	&	\ldots	&	0.908	&	\ldots	&	1.59	&	\ldots	\\
K22	&	-95.5	&	60.1	&	-58	&	112.8	&	4.79	&	-57	&	107.0	&	7.04	&	\ldots	&	\ldots	&	\ldots	&	0.949	&	\ldots	&	-2.25	&	\ldots	\\
K23	&	76.5	&	83.7	&	42	&	113.4	&	4.84	&	\ldots	&	\ldots	&	\ldots	&	\ldots	&	\ldots	&	\ldots	&	\ldots	&	\ldots	&	\ldots	&	\ldots	\\
K24	&	73.3	&	89.9	&	39	&	116.0	&	5.43	&	39	&	105.1	&	5.82	&	38	&	102.3	&	5.51	&	0.906	&	0.882	&	-0.39	&	0.31	\\
K25	&	108.0	&	42.8	&	68	&	116.1	&	5.68	&	68	&	106.0	&	7.89	&	\ldots	&	\ldots	&	\ldots	&	0.913	&	\ldots	&	-2.21	&	\ldots	\\
K26	&	97.3	&	77.4	&	51	&	124.3	&	5.88	&	\ldots	&	\ldots	&	\ldots	&	\ldots	&	\ldots	&	\ldots	&	\ldots	&	\ldots	&	\ldots	&	\ldots	\\
K27	&	124.0	&	-14.9	&	97	&	124.9	&	5.36	&	97	&	114.8	&	6.78	&	97	&	113.1	&	6.30	&	0.919	&	0.906	&	-1.42	&	0.48	\\
K28	&	-110.1	&	61.2	&	-61	&	126.0	&	5.21	&	-61	&	115.9	&	5.79	&	-61	&	114.4	&	5.59	&	0.920	&	0.908	&	-0.58	&	0.20	\\
K29	&	123.3	&	37.6	&	73	&	128.9	&	4.78	&	73	&	116.9	&	7.10	&	73	&	116.0	&	6.66	&	0.907	&	0.900	&	-2.32	&	0.44	\\
K30	&	122.5	&	46.9	&	69	&	131.2	&	5.48	&	70	&	119.5	&	6.79	&	69	&	118.8	&	7.40	&	0.911	&	0.905	&	-1.31	&	-0.61	\\

\enddata
    
\end{deluxetable}

\clearpage
\begin{figure}
\epsscale{0.8}
\plotone{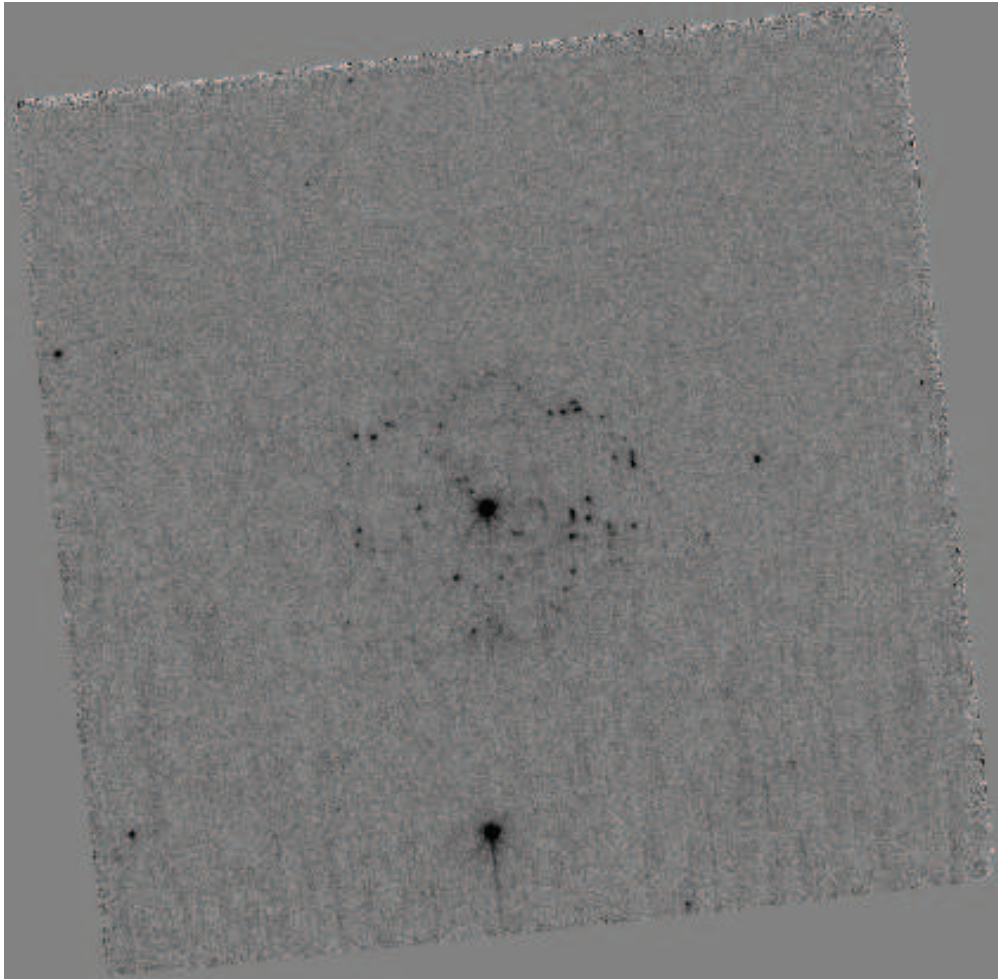}
\caption{
T Pyx field.  This picture shows our full image from 2007 June 29 with the Planetary Camera on the WFPC2.  T Pyx itself is the star near the middle, with the star Check near the bottom, and three faint background stars visible (two near the left edge and one towards the right in the middle).  The nova shell is visible as knots circularly distributed around T Pyx, with 30 knots being relatively bright (and many more visible).  In this image (also Figures 2 and 3), north is directly to the top and east is to the left.  The field size is 800$\times$800 pixels, or 35$\times$35 arc-seconds.}
\end{figure}

\clearpage
\begin{figure}
\epsscale{0.8}
\plotone{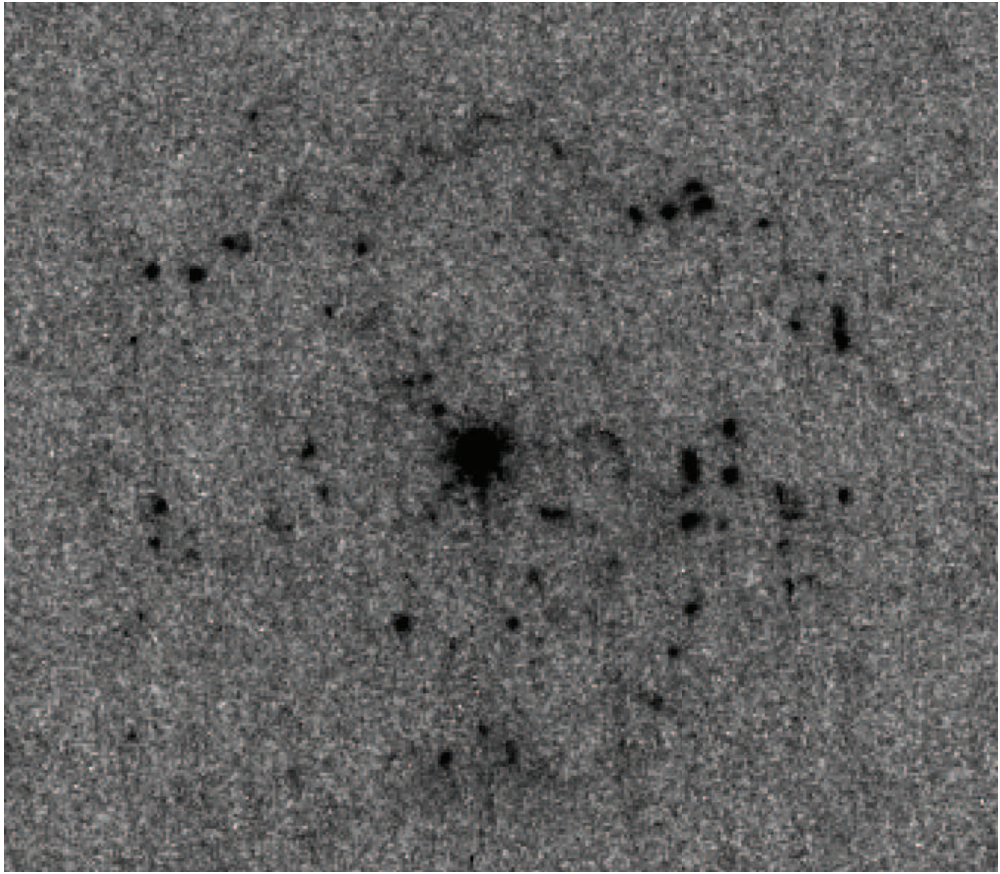}
\caption{
T Pyx knots in 2007.  This picture is a close-up of Figure 1 so that the knots can be better seen, and the individual knots are numbered as in Table 1.  All knots are unresolved even though some knots appear close to other knots.  We see that the knots are mostly at radial distances of 100-130 pixels (4.6-6.0 arc-seconds) from T Pyx.  This distribution points to the shell being hollow with a finite thickness.  This image can be directly compared to Figures 2, 6, and 7 in Shara et al. (1997) of {\it HST} images in 1994 and 1995.  In this comparison, we readily see that many of the knots form identical patterns in 1994, 1995, and 2007, and this provides our assurance that we are seeing the same knots in all three epochs.  We also see some knots that do not appear in the 1995 image, and the turn-on of knots at this late epoch is consistent only with the knots being powered by shocks from collisions between ejecta from different eruptions.}
\end{figure}

\clearpage
\begin{figure}
\epsscale{1.0}
\plotone{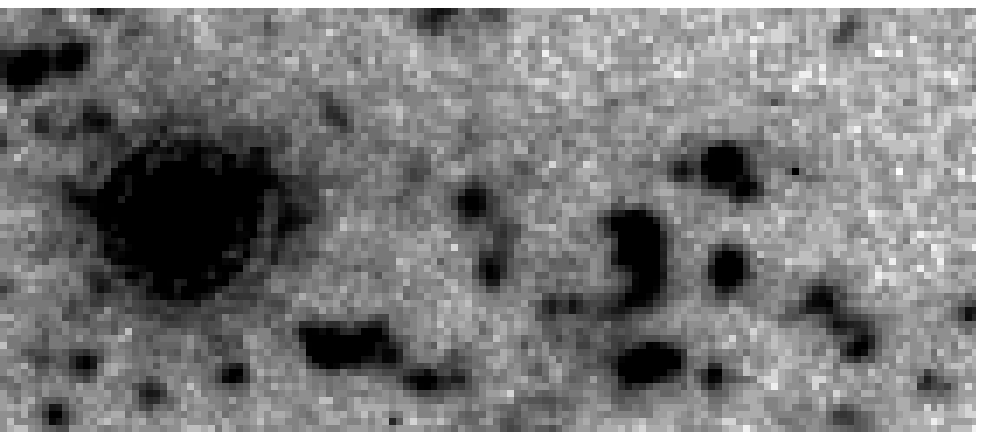}
\plotone{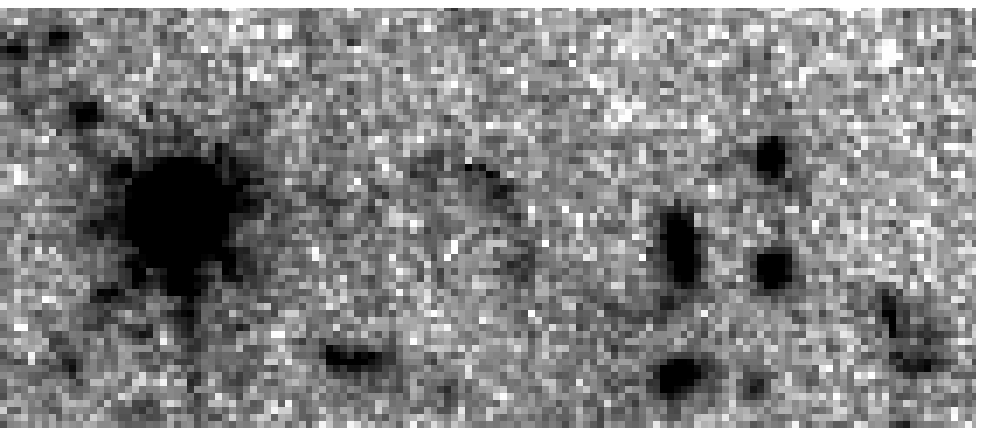}
\caption{
The expansion from 1995 to 2007.  The close-ups from 1995 (top panel) and 2007 (bottom panel) have identical scaling, positioning, and cropping.  The knots in both images form `constellations'  which are distinctly visible in both images, and the duplication of these complex patterns provides confidence that we are seeing the {\it same} knots in both images.  We see that T Pyx has the same position, but that the constellations of knots are shifted to the right by a small but noticeable and significant amount.  This provides visual proof of the expansion of the shell.  We also see that the 1995 image goes substantially deeper than the 2007 image, due to the much greater exposure time in 1995 as compared to our total of 3800 seconds in 2007.}
\end{figure}

\clearpage
\begin{figure}
\epsscale{1.0}
\plotone{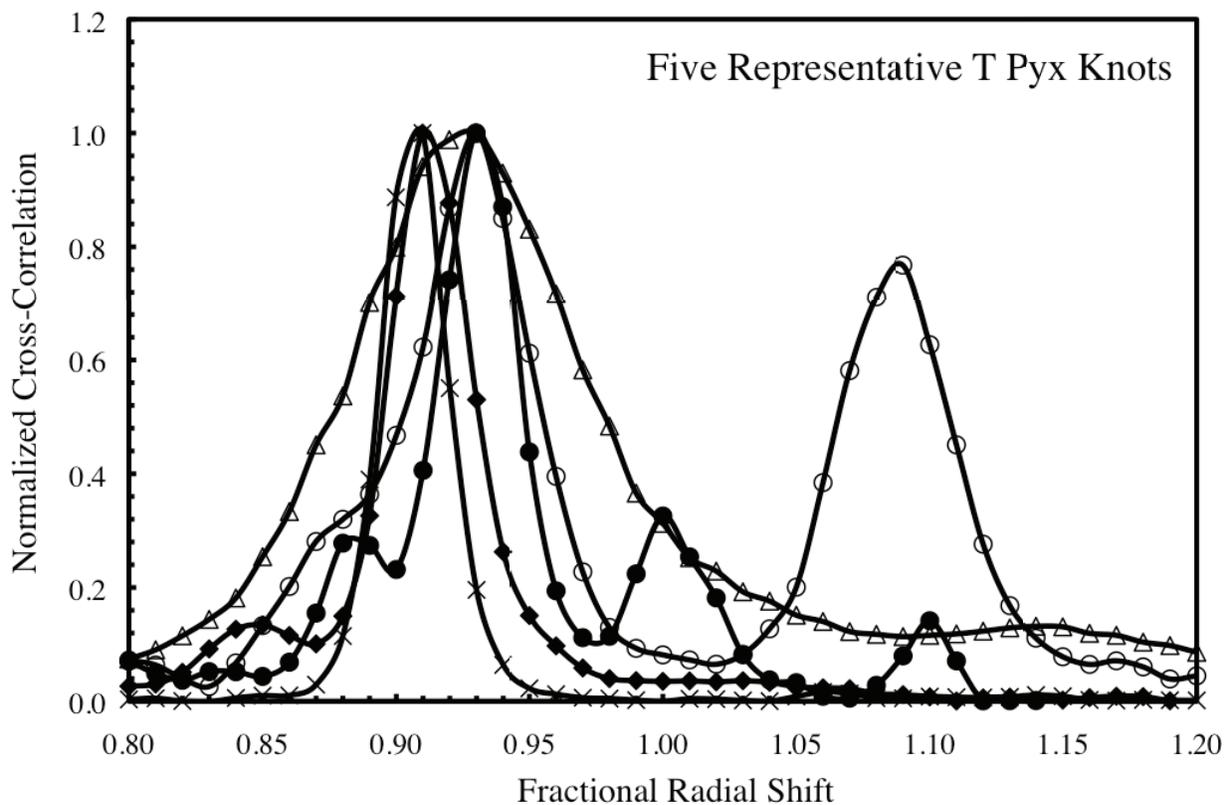}
\caption{
Cross-correlation from 2007 to 1995. The cross-correlation is the product of the flux of a knot in a one arc-second photometric aperture on the 2007 image and the flux in a shifted position on the 1995 image.  The shifts are all radial, where the distance to T Pyx is changed by a factor of $F$, as plotted along the horizontal axis.  This plot gives the cross-correlation for five representative knots.  The peak in the functions shows the fractional radial shift from 2007 to 1995.  All of these five knots (and almost all of the knots) have peaks close to $F=0.91$.  This demonstrates that the choice of 1995 knots to match with the 2007 knots is accurate.  This also demonstrates that the expansion of the T Pyx shell is homologous, in that the radial shifts from 1995 to 2007 are proportional to the radial distance from T Pyx.}
\end{figure}

\clearpage
\begin{figure}
\epsscale{1.0}
\plotone{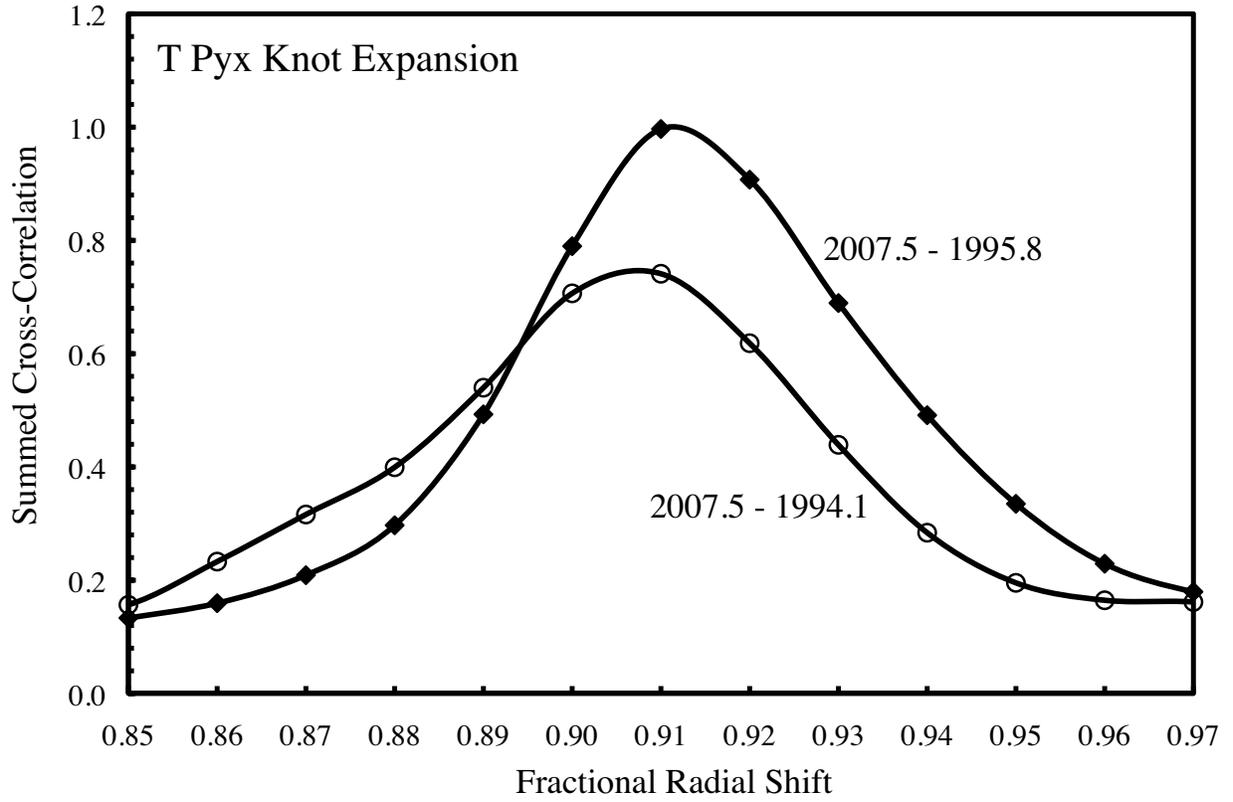}
\caption{
Cross-correlations summed for 30 knots. The summed cross correlations both show a single highly-significant peak near $F=0.91$.  In more detail, we find $F=0.912\pm0.004$ for 2007.5-1995.8 and $F=0.907\pm 0.005$ for 2007.5-1994.1 from this plot.}
\end{figure}

\clearpage
\begin{figure}
\epsscale{1.0}
\plotone{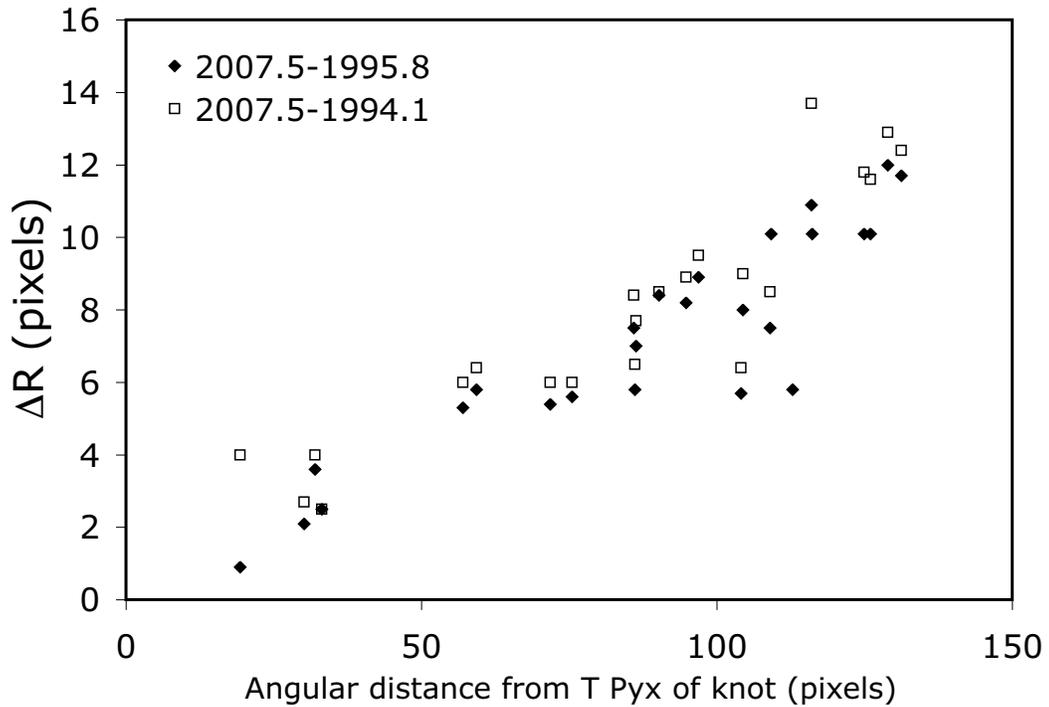}
\caption{
Homologous radial expansion. From eaach of the two images in the middle 1990s to 2007.5, the knots move outward by an amount $\Delta R$ (in pixels).  This expansion is plotted for both 1994.1 to 2007.5 (open squares) and from 1995.8 to 2007.5 (filled diamonds).  We see that the two sets of points form a fairly good line with an intercept close to the origin.  This is the signature of homologous expansion, where a knot twice as far away from T Pyx will move twice as much in a given time interval, and so on.}
\end{figure}

\clearpage
\begin{figure}
\epsscale{0.7}
\plotone{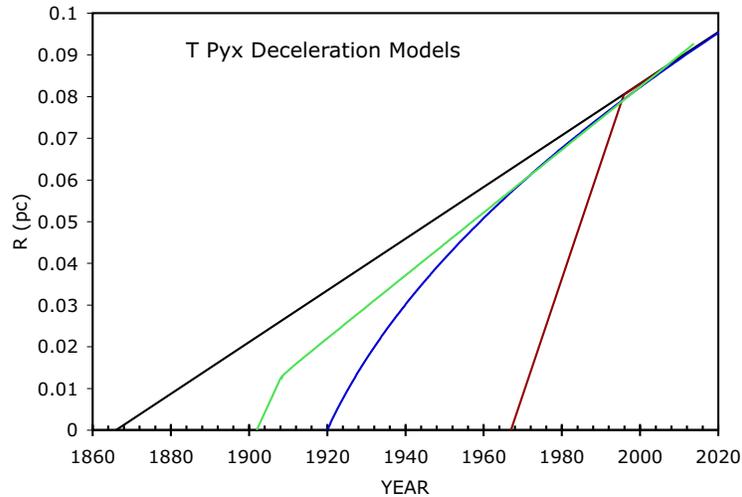}
\caption{
T Pyx deceleration models.  As knots are ejected, their radial distance from T Pyx will increase; with the slope of the line in this plot gives the expansion velocity.  The four displayed curves correspond to the four different models, where the no-deceleration model is shown with an eruption around the year 1866, the sudden deceleration model is shown for the 1967 eruption, the deceleration from the prior shell model is shown for the 1902 eruption, and the deceleration from the ISM model is shown for the 1920 eruption.  None of the models with significant deceleration can work because the degree of deceleration critically depends on the greatly-variable mass of the individual knots, whereas we see all the knots having closely similar fractional expansions.  The only remaining viable model is the one in which the knots have undergone no significant deceleration.  With this, we conclude that the currently-visible knots were ejected around the year 1866 in a classical nova event, and they have been coasting outward ever since.}
\end{figure}

\clearpage
\begin{figure}
\epsscale{1.0}
\plotone{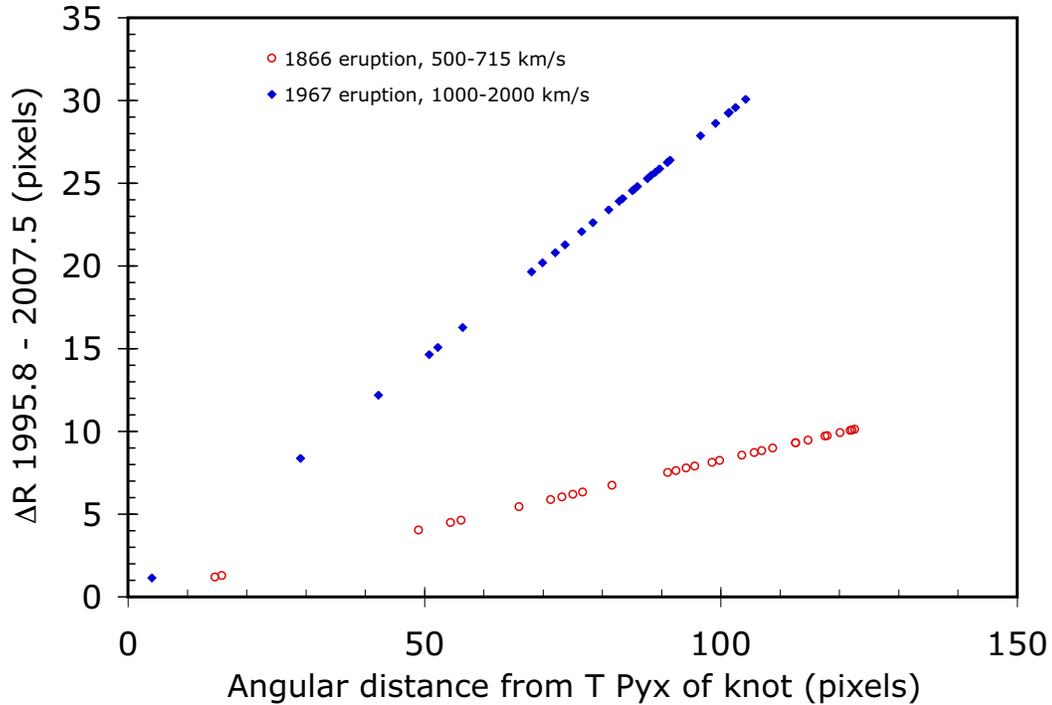}
\caption{
Simulation of T Pyx knots.  This plot of simulated knots shows their change in angular distance from 1995.8 to 2007.5 as a function of the angular distance in 2007.5.  This plot should be directly comparable to Figure 6 for the observed data from T Pyx.  All the knots from the 1866 eruption lie exactly along a straight line pointing to the origin, despite the fact that the knots have widely varying velocities and angles to the line of sight.  Similarly, the knots from the 1967 eruption all fall exactly along a different straight line pointing back to the origin.  The slopes of these lines tell us when the knots were ejected.  This provides a confident means of identifying the eruption date and deciding whether multiple eruptions have co-mingling knots.}
\end{figure}

\clearpage
\begin{figure}
\epsscale{1.0}
\plotone{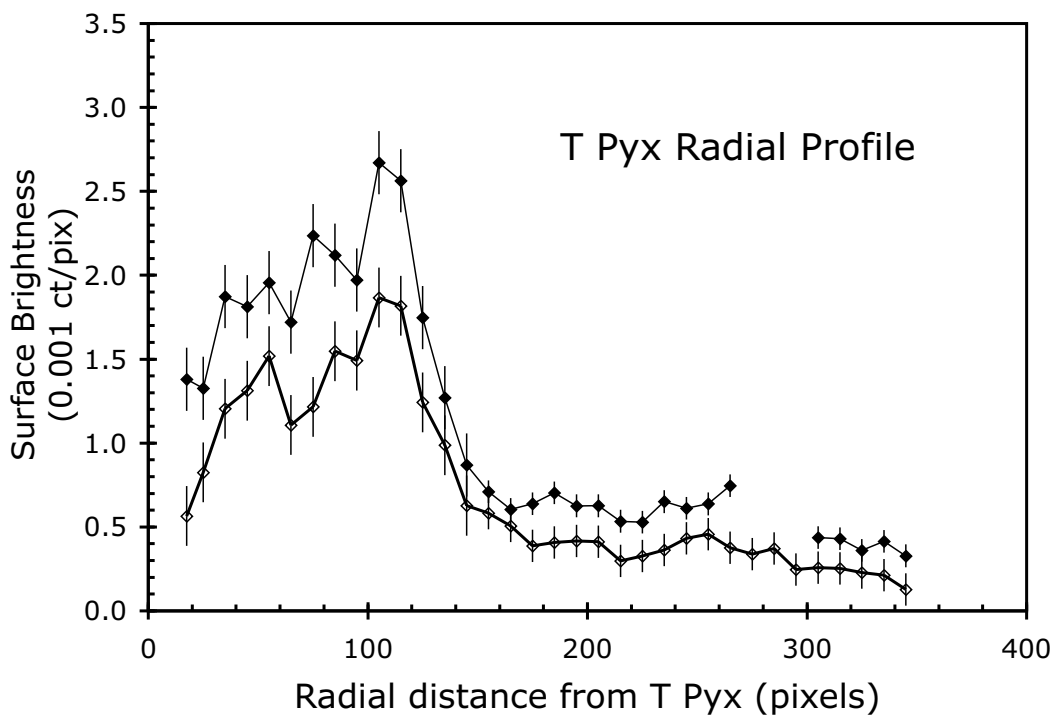}
\caption{
Radial profile of T Pyx.  This radial profile is dominated by the knots from the `1866' eruption out to 150 pixels (with a peak near 100 pixels), with a faint outer halo visible to at least 350 pixels.  The pixel size is 0.045".  The surface brightness is averaged within annular regions with a width of 10 pixels, and has units of 0.001 counts per pixel.  The upper curve gives the total flux (including the knots) within the annular region divided by the area, with a break around a radial distance of 280 pixels due to the Check star being in the annuli.  The lower curve gives the mode of the pixel brightnesses in each of the annuli,  representing the smooth shell brightness with the knots removed.}
\end{figure}

\clearpage
\begin{figure}
\epsscale{1.0}
\plotone{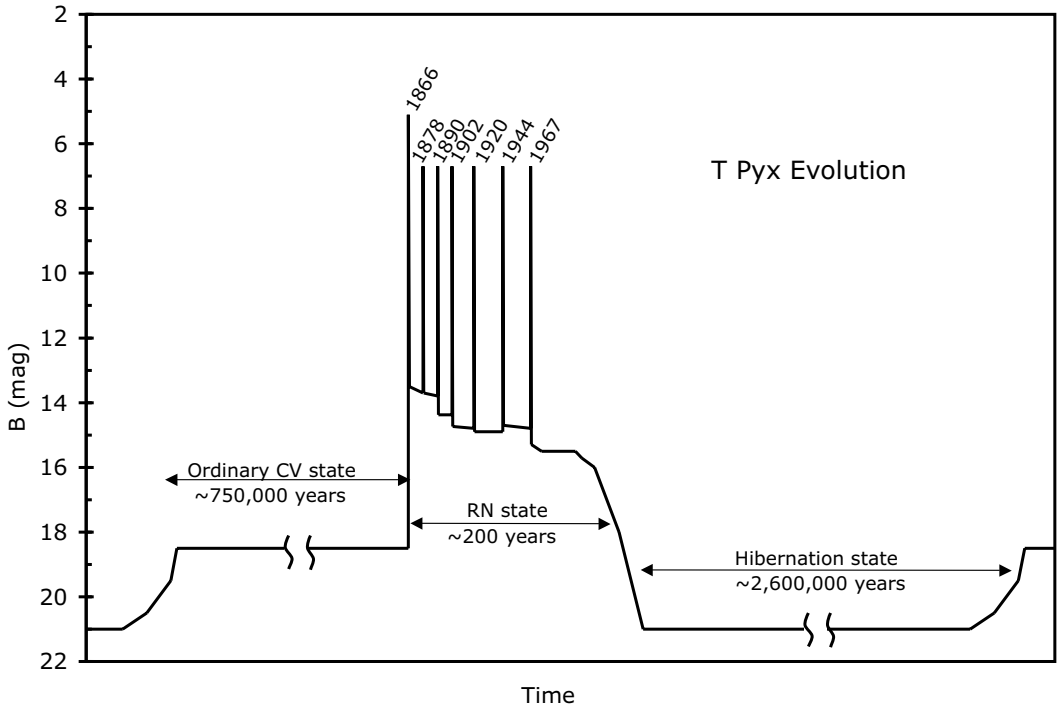}
\caption{
Brightness evolution of T Pyx.  This schematic plot of the B-band brightness of T Pyx through time is useful for visualizing the evolutionary cycle.  The cycle goes between three states: the ordinary CV state, the RN state, and the hibernation state.  The ordinary CV state is where the system appears as an unexceptional interacting binary where the accretion is weak, driven only by gravitational radiation angular momentum loss, as is appropriate for T Pyx being below the period gap.  The RN state starts with the ordinary nova event of `1866', which ejected the now-visible knots in the nova shell and initiated a high mass accretion rate, which makes the system bright and causes RN events.  However, the high accretion rate is not self-sustaining, so the brightness declines and the accretion will soon stop.  The hibernation state begins when the puffed up secondary star contracts inside its Roche lobe and accretion stops.  The hibernation state continues while gravitational radiation slowly brings the secondary star again into contact with its Roche lobe.  As the accretion restarts, the system returns to its ordinary CV state, and the cycle starts over again.}
\end{figure}

\clearpage
\begin{figure}
\epsscale{1.0}
\plotone{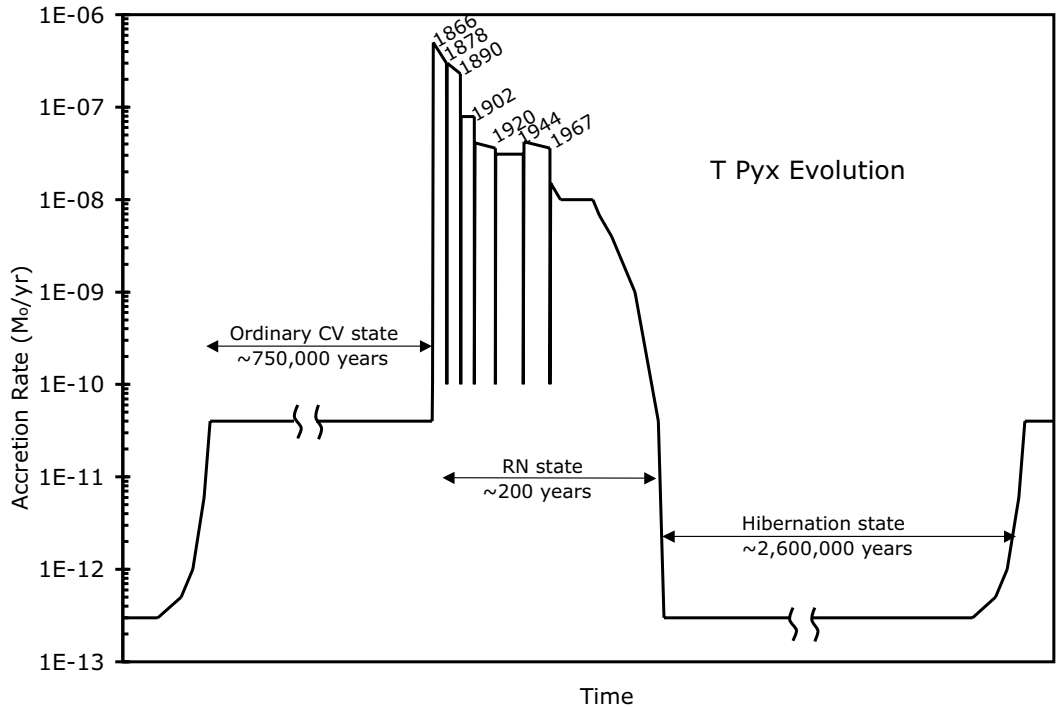}
\caption{
Accretion evolution of T Pyx.  Within our evolutionary scenario for T Pyx, the accretion rate changes in many different ways, so this figure is meant to help visualize these changes.  T Pyx will cycle between three states, with a total cycle time of roughly 3.3 million years.  The accretion during each eruption will briefly fall by some unknown amount due to disruption of the accretion disk by the ejecta.  The end of the hibernation state will have the accretion rising steadily due to material from the outer reaches of the companion's stellar atmosphere crossing the surface of the Roche lobe. The accretion rate steadily declines through the RN state because the supersoft source is not efficient enough to drive a high-enough mass loss to sustain itself; this is the reason why the RN state will inevitably end with T Pyx going into a hibernation state.}
\end{figure}

\end{document}